\documentclass[12pt]{article}%
\usepackage{amsfonts}
\usepackage{setspace}
\usepackage{graphicx}
\usepackage{amsthm}
\usepackage{amsmath}
\usepackage{amssymb}
\usepackage{appendix}
\usepackage{color}
\usepackage{sgame}
\usepackage{verbatim}
\usepackage{caption}
\usepackage{subcaption}
\usepackage{soul}
\usepackage{mathtools}
\usepackage{dsfont}
\usepackage{float}
\usepackage[margin=1in]{geometry}
\usepackage{natbib}
\usepackage[utf8]{inputenc}
\usepackage[english]{babel}
\usepackage{lscape}
\usepackage[colorlinks=true,allcolors=blue]{hyperref}
\usepackage{booktabs}
\usepackage{longtable}
\usepackage{multirow}
\usepackage{threeparttable}
\usepackage{dcolumn}
\usepackage{tikz}
\usepackage{xcolor}
\usepackage{istgame}%
\usetikzlibrary{calc}
\newtheorem{definition}{Definition}

\newtheorem{remark}{Remark}

\newtheorem{proposition}{Proposition}

\DeclareMathOperator*{\argmax}{argmax}
\begin{document}
\begin{singlespace}
\title{A Comparison of Cursed Sequential Equilibrium and Sequential Cursed Equilibrium: Different Concepts of Cursedness in Dynamic 
Games\thanks{This paper was previously circulated under the title ``A Note on Cursed Sequential Equilibrium and Sequential Cursed Equilibrium.'' Support from the National Science Foundation (SES-2243268, SES-2343948) and the Qualitative Collaboration at the University of Virginia is gratefully acknowledged. We thank S. Nageeb Ali, Alexander L. Brown, Paul Cheung, Shani Cohen, Tony Q. Fan, Menglong Guan, Paul J. Healy, Shengwu Li, Emanuel Vespa, Kai Hao Yang, and Junya Zhou for helpful discussions and comments.}}
\author{Meng-Jhang Fong\thanks{%
Max Planck Institute for Behavioral Economics, Bonn, Germany 53113. fong@econ.mpg.de} \and Po-Hsuan Lin%
\thanks{
University of Virginia, Charlottesville, VA 22904 USA. plin@virginia.edu} \and Thomas R. Palfrey%
\thanks{%
California Institute of Technology, Pasadena, CA 91125 USA.
trp@hss.caltech.edu.} }
\date{\today}
\maketitle
\begin{abstract}
Cursed Equilibrium of \cite{eyster2005cursed} has been a leading theory for explaining winner’s-curse-type behavior in static Bayesian games, but it faces conceptual limitations when applied to dynamic games.
Two recent extensions, Cursed Sequential Equilibrium (CSE) by 
\cite{fong2023cursed} and Sequential Cursed Equilibrium (SCE) by 
\cite{cohen2022sequential}, 
address these limitations in fundamentally different ways.
Complementing these two papers, this paper 
provides a systematic comparison of CSE and SCE, clarifying their conceptual foundations and technical implications, including their notions of cursedness, belief updating, treatment of public histories, and relationship to cursed equilibrium in one-stage Bayesian games.

\end{abstract}

\bigskip

\small{

}
\end{singlespace}

\thispagestyle{empty}

\newpage\setcounter{page}{1}

\section{Introduction}

\label{sec:intro}

Since the seminal work of \cite{eyster2005cursed}, cursed equilibrium (CE) has emerged as a powerful solution concept for explaining behavior that standard Bayesian Nash equilibrium fails to capture in simultaneous-move games with incomplete information. This behavioral equilibrium concept formalizes the
idea that individuals fail to fully account for the correlation between the actions and private types of
other players, with the leading example being the winner's curse in common-value auctions. Thus, a fully cursed player best responds to the average distribution of actions of other players, completely ignoring the dependence of other players' actions on their types, and a CE is a fixed point of these cursed best responses. In a partially cursed equilibrium ($\chi$-CE), players best respond to beliefs that partially take account of the dependence of actions on types, where $\chi$ denotes the weight on fully cursed beliefs, with $\chi=0$ corresponding to Bayesian Nash Equilibrium. 

While CE (and $\chi$-CE) has proved useful because of its tractability and explanatory power for anomalous behavior such as the winner’s curse, its formulation as a strategic-form equilibrium concept means that CE is restricted to being the same in dynamic games with different timing but identical strategic forms, which 
leads to significant limitations when applied to dynamic games.
Two recent papers propose different approaches to modeling cursedness in dynamic games: the Cursed Sequential Equilibrium (CSE) of \cite{fong2023cursed} and the Sequential Cursed Equilibrium (SCE) of \cite{cohen2022sequential}. 
While both approaches build on CE, they formalize cursed behavior in fundamentally different ways and are developed in different formal frameworks. 
As a result, their underlying assumptions and implications are not straightforward to compare, making it difficult for researchers to understand how the two solution concepts relate to each other and how they should be applied.
This paper provides a direct comparison of CSE and SCE in multistage games with observed actions, a class of games that is common in applications and to which both solution concepts apply. 
Within this unified setting, we use illustrative examples and clear, intuitive derivations to highlight the similarities and differences between the two concepts, clarify their respective implications, and facilitate their application to dynamic games.

CSE extends the notion of cursed behavior to
multistage games with observed actions by assuming that, after
each stage of the game, each player in the game updates their current beliefs
about the profile of other players' types in a fully cursed way, i.e., based on the
\emph{average behavioral strategy} of the other players rather than the type-conditional behavioral strategy of the other players. Players are assumed to be sequentially rational, but with best responses relative to this kind of cursed updating about the distribution of types of the other players and to the implied average behavioral strategies of others in future stages. As in CE, CSE is generalized to $\chi$-CSE, where the parameter $\chi\in[0,1]$ measures the
degree to which the correlation between actions and types is ignored, with $\chi=0$ corresponding to standard sequential equilibrium
(i.e., Bayesian updating consistent with the type-conditional 
behavioral strategies of the other players) and $\chi=1$ to the \emph{fully
cursed} case. Values of $\chi\in(0,1)$ correspond to a mixture of these
two extremes, defined similarly to CE, but with respect to behavioral
strategies rather than mixed strategies.

SCE extends the analysis of cursed behavior to games in extensive form with perfect recall by defining a player's cursedness in terms of coarsening the partition of other players' information sets. By coarsening other players’ information sets, SCE captures a player’s failure to infer how other players’ unobserved or hypothetical actions may depend on the history of play, which is not reflected in CSE. At the same time, SCE preserves correct Bayesian inference about types from observed past actions, in contrast to CSE, so the SCE concept of cursed behavior is based on distorted beliefs about unobserved (simultaneous or future) actions. Like CE and CSE, SCE is generalized to $\phi$-SCE,\footnote{To avoid potential confusion, we use $\phi$ to denote
the cursedness parameter for SCE, which corresponds to $\chi$ in \cite{cohen2022sequential}.} where the parameter $\phi\in[0,1]$ scales the degree to which players neglect information from unobserved actions. In the limiting case with $\phi=0$, $\phi$-SCE coincides with standard sequential equilibrium similarly to the boundary case of $\chi$-CSE. However, for $\phi>0$ and $\chi>0$, the solutions are typically different, as we explain in the remainder of the paper.\footnote{The difference between SCE and CSE is reminiscent of the differences between CE and the Analogy-Based Expectations Equilibrium (ABEE) approach by \cite{jehiel2005analogy} and \cite{jehiel2008revisiting}, where the latter approach is based on the bundling of nodes at which other players move into analogy classes, which formally is a (partial) coarsening of information sets, while CE is developed for applications to Bayesian games in the strategic form which correspond to one-stage games in CSE.}


From the perspective of experimental research, it is fortunate that two distinct extensions of CE to dynamic games are now available, because each organizes a different set of laboratory findings that sequential equilibrium cannot account for. \cite{fong2023cursed} show that CSE is consistent with experiments in which subjects appear to draw too little inference from observed actions about others’ private information, including signaling games \citep{brandts1993adjustment}, reputation building \citep{jung1994existence}, durable-goods monopoly \citep{guth1995durable}, the dirty faces game \citep{lin2022cognitivemultii}, information cascades \citep{goeree2007self}, and cheap talk \citep{cai2006overcommunication}. \cite{cohen2022sequential}, in turn, show that SCE is consistent with experiments in which subjects draw better inferences from realized than from hypothetical events, including pivotal voting \citep{esponda2014hypothetical}, learning from prices \citep{ngangoue2021learning}, bid revision in common-value auctions \citep{moser2019hypothetical}, and, through an adaptation to clock auctions, dynamic versus static auctions \citep*{levin1996revenue, levin2016separating}.

These two bodies of evidence suggest when each concept may be particularly relevant. 
CSE may be the more natural candidate when the regularity of interest is dampened updating about others’ private information from observed play, as in reputation building, intertemporal pricing, and social learning, including private-value environments where CE has no bite. 
SCE may be the more natural candidate when subjects draw correct inferences once an event is observed but fail to condition on it in advance, or when cursedness concerns endogenous rather than exogenous information. This is not a horse race: neither concept refines the other, in many games both depart from sequential equilibrium in similar directions, and the same experimental manipulation, such as moving from simultaneous to sequential play, need not separate them. Rather, by confronting the same data with both concepts, researchers can learn which form of cursedness is at work. We are hopeful that this paper can guide future researchers who seek to design experiments that test the predictions of CSE and SCE against each other, or that identify different forms of cursedness within the same setting.

The remainder of the paper is organized as follows. 
Section \ref{sec:prelim} introduces the framework and the two solution concepts. 
Section \ref{sec:differences} examines four aspects in which CSE and SCE differ: their underlying notions of cursedness (Section \ref{subsec:cursedness_notion}), their implications for belief-updating dynamics (Section \ref{subsec:belief_updating_new}), their treatment of public histories (Section \ref{subsec:public_history}), and their predictions in one-stage simultaneous-move games (Section \ref{subsec:one_stage}).
Finally, Section \ref{sec:conclusion} concludes the paper.

\section{Notation and Definitions}
\label{sec:prelim}


\subsection{Multistage Games with Observed Actions}
\label{subsec:multistage}

In this paper, we focus on the differences between the two solution concepts within the framework of multistage games with observed actions \citep{fudenberg1991perfect}. A \textit{multistage game with observed actions}, $\Gamma$, is defined by the following components:

\begin{itemize}
    \item A finite set of \textit{players}, $N = \{1, ..., n \}$.
    \item A finite set of \textit{type profiles}, $\Theta\equiv\times_{i=1}^{n}\Theta_{i}$, where $\theta_{i}\in\Theta_{i}$ denotes player $i$'s type, which is private information.
    \item A full-support common \textit{prior distribution of type profiles}, $F:\Theta\rightarrow(0,1)$.
    \item A sequence of \textit{stages}, $t=1,2,\ldots,T$.
    \item A set of \textit{public histories} of past action profiles, $\mathcal{H}=\bigcup_{t=0}^{T}\mathcal{H}^{t}$, where $\mathcal{H}^{t-1}$ denotes the set of all possible public histories at stage $t$, with $\mathcal{H}^{0}=\{h_{\emptyset}\}$ denoting the null public history in stage 1, $\mathcal{H}^{T}$ the set of terminal public histories, and $\mathcal{H}\backslash\mathcal{H}^{T}$ the set of non-terminal public histories. We define a partial order $\prec$ on $\mathcal{H}$ such that $h\prec h^{\prime}$ means $h$ precedes $h^{\prime}$.

    \item An \emph{information set} of player $i$ consists of a non-terminal public history and player $i$'s own type, denoted by $I_i = (h^{t-1}, \theta_i)$ with $h^{t-1} \in \mathcal{H}\backslash \mathcal{H}^T$ and $\theta_i \in \Theta_i$. Let $\mathcal{I}_i \equiv \mathcal{H} \backslash \mathcal{H}^T \times \Theta_i$ denote the set of player $i$'s information sets, and let $\mathcal{I} \equiv \bigcup_{i \in N} \mathcal{I}_i$ denote the set of all information sets.
    
    \item A type-independent finite set of \textit{feasible actions} for each player $i$, $A_{i}=\bigcup_{h\in\mathcal{H}\backslash\mathcal{H}^{T}}A_{i}(h)$, where $|A_{i}(h)|\geq1$ for all $i\in
N$ and any $h\in\mathcal{H}\backslash\mathcal{H}^{T}$.
    \item A \textit{payoff function} for each $i$, $u_{i}:\mathcal{H}^{T} \times \Theta \rightarrow\mathbb{R}$.    
    \item A \textit{behavioral strategy} for each player $i$, $\sigma_{i}:\mathcal{I}_{i}\rightarrow\Delta(A_{i})$, with $\sigma_{i}( h^{t-1}, \theta_i)\in\Delta(A_{i}(h^{t-1}))$. Furthermore, let $\sigma_{i}(a_{i}^{t}|h^{t-1}, \theta_{i})$ denote the probability player $i$ chooses $a_{i}^{t}\in A_{i}(h^{t-1})$.
\end{itemize}

\subsection{Cursed Sequential Equilibrium in Multistage Games}
\label{subsec:cse_notations}

An \textit{assessment} is denoted by $(\mu,\sigma)$, where $\mu$ is a belief system and
$\sigma$ is a behavioral strategy profile. The belief system specifies, for each type of
each player, a conditional distribution over the set of type profiles of other players
at each public history. Following the spirit of the cursed
equilibrium, for type $\theta_{i}$ player $i$ at public history $h^{t-1}$, we define the
\emph{average behavioral strategy profile of the other players} as:
\[
\bar{\sigma}_{-i}(a_{-i}^{t}|h^{t-1}, \theta_{i})=\sum_{\theta_{-i}\in
\Theta_{-i}}\mu(\theta_{-i}|h^{t-1}, \theta_{i})\sigma_{-i}(a_{-i}%
^{t}|h^{t-1}, \theta_{-i})
\]
for any $i\in N$, $\theta_{i} \in\Theta_{i} $ and $h^{t-1}\in\mathcal{H}%
^{t-1}$.

A $\chi$-CSE is parameterized by a single parameter $\chi\in[0,1]$. Instead of
thinking the other players are using $\sigma_{-i}$, a $\chi$-cursed type
$\theta_{i}$ player $i$ would believe they are using a $\chi$-weighted average
of the average behavioral strategy and the true behavioral strategy:
\[
\sigma_{-i}^{\chi}(a_{-i}^{t}|h^{t-1}, \theta_{-i},\theta_{i})=\chi\bar{\sigma
}_{-i}(a_{-i}^{t}|h^{t-1}, \theta_{i})+(1-\chi)\sigma_{-i}(a_{-i}^{t}%
|h^{t-1}, \theta_{-i}).
\]

The beliefs of player $i$ about the type profile $\theta_{-i}$ are updated via
Bayes' rule in $\chi$-CSE, whenever possible, assuming others are using the
$\chi$-cursed behavioral strategy rather than the true behavioral strategy.
Specifically, an assessment satisfies the \textit{$\chi$-cursed Bayes' rule}
if the belief system is derived from Bayes' rule while perceiving others are
using $\sigma_{-i}^{\chi}$ rather than $\sigma_{-i}$.

\begin{definition}
[$\chi$-cursed Bayes' rule, Definition 1 of \citealp{fong2023cursed}]%
\label{def:cursed_bayes} An assessment $(\mu,\sigma)$ satisfies $\chi$-cursed
Bayes' rule if the following rule is applied to update the posterior beliefs
whenever $\sum_{\theta_{-i}^{\prime}\in\Theta_{-i}}\mu(\theta_{-i}%
^{\prime}|h^{t-1}, \theta_{i})\sigma_{-i}^{\chi}(a_{-i}^{t}|h^{t-1}, \theta_{-i}%
^{\prime},\theta_{i})>0$:
\[
\mu(\theta_{-i}|h^{t}, \theta_{i})=\frac{\mu(\theta_{-i}|h^{t-1}, \theta
_{i})\sigma_{-i}^{\chi}(a_{-i}^{t}|h^{t-1}, \theta_{-i},\theta_{i}%
)}{\sum_{\theta_{-i}^{\prime}\in\Theta_{-i}}\mu(\theta_{-i}^{\prime
}|h^{t-1}, \theta_{i})\sigma_{-i}^{\chi}(a_{-i}^{t}|h^{t-1}, \theta_{-i}^{\prime}%
,\theta_{i})}.
\]

\end{definition}

Finally, $\chi$-CSE places a consistency restriction, analogous to consistent
assessments in sequential equilibrium, on how $\chi$-cursed beliefs are
updated off the equilibrium path.

\begin{definition}
[$\chi$-consistency, Definition 2 of \citealp{fong2023cursed}]%
\label{def:chi_consistency} Let $\Psi^{\chi}$ be the set of assessments
$(\mu,\sigma)$ such that $\sigma$ is a totally mixed behavioral strategy
profile and $\mu$ is derived from $\sigma$ using $\chi$-cursed Bayes' rule.
$(\mu,\sigma)$ satisfies $\chi$-consistency if there is a sequence of
assessments $\{(\mu^{k},\sigma^{k})\}\subseteq\Psi^{\chi}$ such that
$\lim_{k\rightarrow\infty}(\mu^{k},\sigma^{k})=(\mu,\sigma)$.
\end{definition}

For any $i \in N$, $\chi\in[0,1]$, $\sigma$, and $\theta\in\Theta$, let
$\rho_{i}^{\chi}(h^{T}|h^{t}, \theta,\sigma_{-i}^{\chi},\sigma_{i})$ be player
$i$'s perceived conditional realization probability of terminal history $h^{T}
\in\mathcal{H}^{T}$ at the public history $h^{t} \in\mathcal{H}\backslash
\mathcal{H}^{T}$ if the type profile is $\theta$ and player $i$ uses the
behavioral strategy $\sigma_{i}$ whereas perceives other players' using the
cursed behavioral strategy $\sigma_{-i}^{\chi}$. At every non-terminal public
history $h^{t}$, a $\chi$-cursed player in $\chi$-CSE will use $\chi$-cursed
Bayes' rule (Definition \ref{def:cursed_bayes}) to derive the posterior belief
about the other players' types. Accordingly, a type $\theta_{i}$ player $i$'s
conditional expected payoff at the public history $h^{t}$ is given by:
\[
\mathbb{E}u^\chi_{i}(\sigma|h^{t}, \theta_{i}) = \sum_{\theta_{-i} \in\Theta_{-i}}
\sum_{h^{T} \in\mathcal{H}^{T}} \mu(\theta_{-i}|h^{t}, \theta_{i}) \rho
_{i}^{\chi}(h^{T}|h^{t}, \theta,\sigma_{-i}^{\chi},\sigma_{i}) u_{i}%
(h^{T}, \theta_{-i}, \theta_{i}).
\]

\begin{definition}
[$\chi$-cursed sequential equilibrium, Definition 3 of
\citealp{fong2023cursed}]An assessment $(\mu^{*}, \sigma^{*})$ is a $\chi
$-cursed sequential equilibrium if it satisfies $\chi$-consistency and
$\sigma_{i}^{*}(h^{t}, \theta_{i})$ maximizes $\mathbb{E}u^\chi_{i}(\sigma
^{*}|h^{t}, \theta_{i})$ for all $i$, $\theta_{i}$, $h^{t} \in\mathcal{H}%
\backslash\mathcal{H}^{T}$.

\end{definition}

\subsection{Sequential Cursed Equilibrium in Multistage Games}
\label{subsec:sce_notations}

Some additional notation is required to define sequential cursed equilibrium
(SCE) in multistage games with observed actions. First, let $\lambda$ denote nature who
only moves once in stage~0, i.e., at the initial history, denoted $h_{\emptyset}$.
That is, nature's information set is singleton. The
action set for $\lambda$ is the set of type profiles
$\Theta=\times_{i=1}^{n}\Theta_{i}$. Because nature is explicitly modeled as a player in the game, for any multistage game with observed actions, a \textit{history} of the game at any point in time includes both the public history, $h^t$ and nature's move, $\theta$. The totally
mixed strategy of nature, denoted by $\sigma_{\lambda}(h_{\emptyset})\in
\Delta(\Theta)$, corresponds to the prior distribution of types 
$F$, which is common knowledge among all players. 

Second, for any multistage game with observed actions, let 
$\mathcal{P}$ denote the \textit{coarsest valid partition}\footnote{A partition of the non-terminal histories is invalid if it defines information sets that violate the assumption of perfect recall.} of the
non-terminal histories $\mathcal{H}\backslash\mathcal{H}^{T} \times 
\Theta$, and
$P(I_i)$ denotes $P\in\mathcal{P}$ such that $I_i\subseteq P$. 
Note that a partition $\mathcal{P}^{\prime}$ is coarser than
$\mathcal{P}^{\prime\prime}$ if, given any $P^{\prime\prime}\in\mathcal{P}%
^{\prime\prime}$, there exists a cell of $\mathcal{P}^{\prime}$ that contains
$P^{\prime\prime}$.
With this definition, we say that a strategy $\sigma_i$ is \emph{coarse} if 
it is measurable with respect to $\mathcal{P}$.


Third, for any information set $I$, we use $\mathcal{Q}_i^I$ to denote the
collection of player $i$’s information sets that are \emph{compatible}
with $I$.\footnote{In the framework of multistage games with observed
actions, if the information set $I = (h^t, \theta_i)$ belongs to player
$i$, then
$\mathcal{Q}_{i}^{I}
= \{(h, \theta_i) \in \mathcal{I}_i : h \in \mathcal{H} \backslash \mathcal{H}^{T},
\; h^{t} \preceq h \;\mbox{or}\; h \preceq h^t \}.$
On the other hand, if the information set $I = (h^t, \theta_j)$ belongs to
another player $j \neq i$, then
$\mathcal{Q}_{i}^{I}
= \{(h, \theta_i') \in \mathcal{I}_i : h \in \mathcal{H} \backslash \mathcal{H}^{T},
\; h^{t} \preceq h \;\mbox{or}\; h \preceq h^t,\; 
\theta_i' \in \Theta_i\}.$
Thus, in the framework of multistage games with observed actions, a
necessary condition for two information sets to be compatible is that
they contain compatible public histories.}
Furthermore, we use
$\sigma_i^I : \mathcal{Q}_i^I \rightarrow \Delta(A_i)$
to denote a \emph{partial strategy of player $i$ at $I$}, namely, player
$i$’s behavioral strategy with its domain restricted to the information
sets that are compatible with $I$.

Fourth, nature’s mixed strategy is included in the strategy profile. That is, we denote
$\sigma \equiv (\sigma_{1}, \ldots, \sigma_{n}, \sigma_{\lambda}).$
We let $\nu[\sigma]$ denote the probability measure on
$\mathcal{H}^{T} \times \Theta$ induced by $\sigma$.\footnote{To avoid any potential confusion,
throughout this paper we use $\mu$ to denote the belief system of CSE and $\nu$
to denote the induced probability measure of SCE.} In other words,
$\nu[\sigma](h^{T}, \theta)$ is the probability of reaching the terminal
history $(h^{T}, \theta)$ under the strategy profile $\sigma$.
Following \cite{cohen2022sequential}, we extend $\nu[\sigma]$ to non-terminal
histories by viewing each such history as the collection of terminal histories
that succeed it. For any $h \in \mathcal{H} \backslash \mathcal{H}^{T}$ and any
$\theta \in \Theta$,
$\nu[\sigma](h, \theta) \equiv
\sum_{h^{T} \in \{h' \in \mathcal{H}^{T} : h \prec h'\}}
\nu[\sigma](h^{T}, \theta).$



Similar to CSE, an SCE is defined as an \emph{assessment}.
One important difference, however, is that an SCE assessment consists of a behavioral 
strategy profile (including nature) and a \emph{system of conjectures}, rather than a belief system as in CSE. 
Specifically, for any information set $I$, a \emph{conjecture} is a profile of 
partial strategies $\bar{\sigma}^I \equiv (\sigma_i^I)_{i \in N \cup \{\lambda\}}$
such that $\nu[\bar{\sigma}^I](I)> 0$. A \emph{system of conjectures} then 
specifies a conjecture at each information set, denoted by
$\bar{\sigma}^{\mathcal{I}}\equiv((\bar{\sigma}^{I_i})_{I_i \in \mathcal{I}i})_{i \in N}$. An SCE assessment 
is \emph{locally rational} if, at each information set, the active player 
best responds to their conjectures.\footnote{See \cite{cohen2022sequential} for the formal 
definition and for further discussion of local rationality.}

In SCE,
an assessment is called \emph{cursed-plausible} if, for any player at any non-terminal
information set, the player has a correct perception of past events---placing probability
one on actions that do not rule out reaching the current information set---while holding
incorrect perceptions about future events, namely believing that all other players
play the same behavioral strategy within a coarse set. Formally, given a totally 
mixed behavioral strategy profile $\sigma$, any non-terminal information
set $I_i = (h^{t-1}, \theta_i)$ belonging to player $i$, and any non-terminal information set
$I_j = (h^{t'-1}, \theta_j)$ of player $j \neq i$, the cursed conjecture about player
$j$ at $I_j$ is given by:
\[
\bar{\sigma}_j^{I_i}(a_j^{t'} | I_j) \equiv
\nu[\sigma]\big(\{((h,\theta), a_j^{t'}) : (h,\theta) \in P(I_j)\}
| I_i \cap P(I_j)\big),
\]
where $a_j^{t'} \in A_j(h^{t'-1})$, and $((h,\theta), a_j^{t'})$ is shorthand for the set
of histories that succeed $(h,\theta)$ and in which player $j$ chooses $a_j^{t'}$ at
stage $t'$.

For any $\phi \in [0,1]$, a
$\phi$-SCE
is defined as an assessment in which, at every non-terminal information set,
a $\phi$-cursed player evaluates their continuation payoff as a $\phi$-weighted average of the expected payoffs under the cursed-plausible conjecture and under correct beliefs about other players' types and behavior, where $\phi$ represents the degree of cursedness in the SCE framework. Based on these evaluations, a $\phi$-SCE is defined
as follows:

\begin{definition}
[$\phi$-sequential cursed equilibrium, \citealp{cohen2022sequential}] 
An assessment $(\sigma,\bar{\sigma}^{\mathcal{I}})$
is a $\phi$-sequential cursed equilibrium if it is
cursed-consistent\footnote{Similar to sequential equilibrium \citep{krepswilson1982}, SCE imposes a consistency requirement: an assessment is cursed-consistent if it is the limit of a sequence of cursed-plausible assessments with totally mixed behavioral strategy profiles. For $\phi$-SCE, the same sequence is also used to derive the Bayesian posterior beliefs. See \cite{cohen2022sequential} for further details.} and, at every information set $I_i$
of each player $i$, the self-conjecture
$\bar{\sigma}_i^{I_i}$ agrees with $\sigma_i$ at $I_i$
and maximizes $i$'s weighted average of two
conditional expected payoffs:
\begin{enumerate}
    \item With weight $\phi$, the expected payoff is evaluated
    under the cursed-plausible conjecture $\bar{\sigma}^{I_i}$.

    \item With weight $1-\phi$, the expected payoff is evaluated
    under the Bayesian conjecture, with other players
    following their actual strategies $\sigma_{-i}$.
\end{enumerate}

\end{definition}

\section{Differences Between CSE and SCE}
\label{sec:differences}

\subsection{Different Notions of Cursedness}
\label{subsec:cursedness_notion}

Building on the standard cursed equilibrium (CE) for simultaneous-move Bayesian games, CSE and SCE adopt different approaches in dynamic settings, reflecting distinct notions of cursedness. As in standard CE, cursedness in the CSE framework captures a player’s neglect of how opponents’ actions depend on their exogenous private types. By contrast, under SCE, cursedness arises from a player’s failure to recognize how other players’ unobserved (future or simultaneous) actions depend on the history of the game, while still making correct inferences from past observed actions.

We use a crisis-bargaining game as our leading example to illustrate the distinct notions of cursedness in CSE and SCE. We also revisit this example later in Section \ref{subsec:public_history} to show how CSE and SCE are solved and to highlight other differences between the two solution concepts. Crisis-bargaining games are widely applied models used in the study of international conflict, with many variations explored in the literature and varying interpretations of actions and payoffs \citep{fearon1994domestic, fearon1994signaling, Morrow1989,signorino1999,smith1999testing}.\footnote{The specific extensive form we analyze here is a simplified version of the games studied in \cite{Lewis2003signaling} and \cite{fearon1994signaling}.}

\begin{figure}[htbp!]
    \centering
    \includegraphics[width=1.0\linewidth]{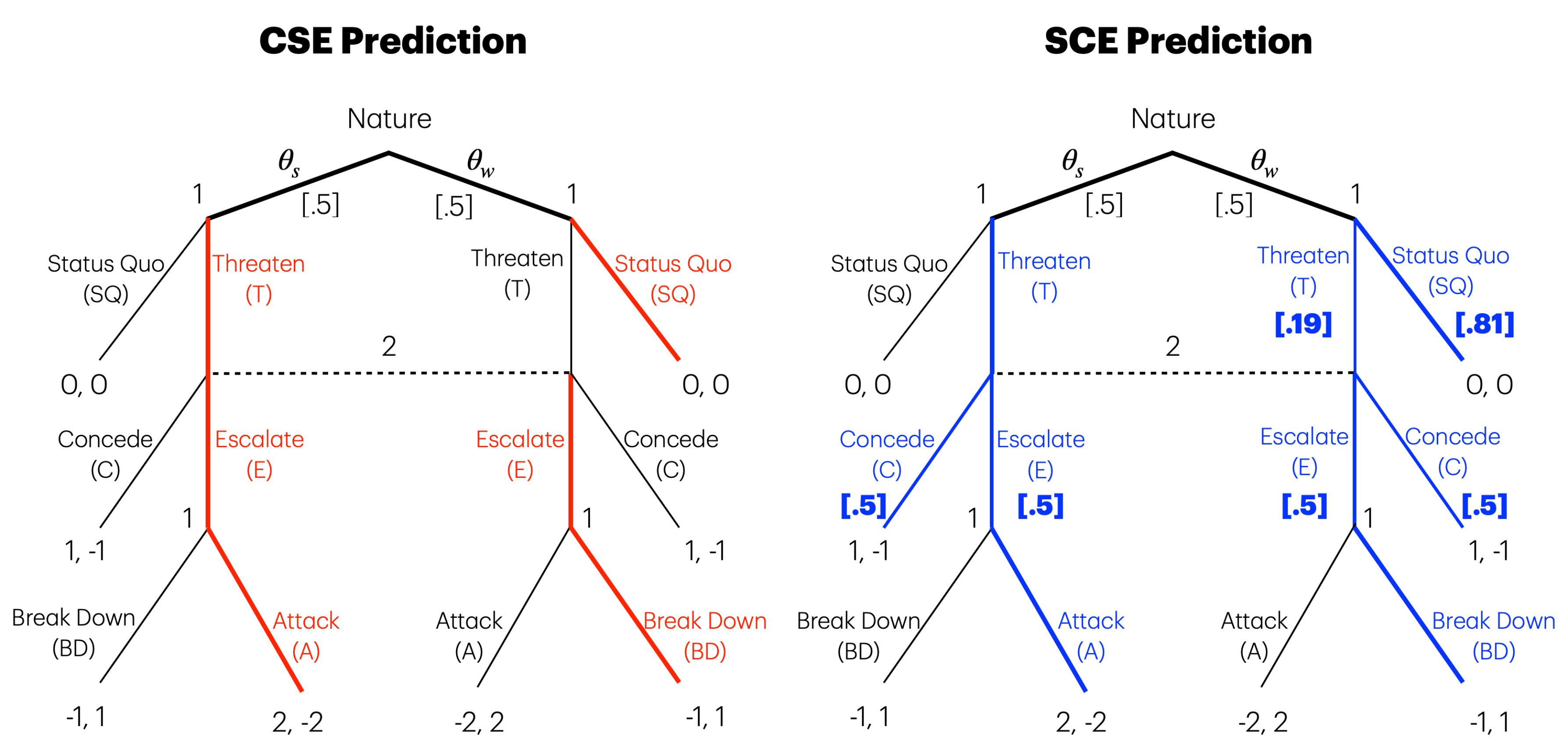}
    \caption{Crisis-bargaining game tree with CSE (left) and SCE (right) solutions.}
    \label{fig:crisis_bargaining_example}
\end{figure}

\bigskip

\noindent\textbf{Illustrative Example 1 (Crisis-Bargaining Game).} Figure \ref{fig:crisis_bargaining_example} presents the game tree of a dynamic crisis-bargaining game with two players. Player 1 may be either strong ($\theta_s$) or weak ($\theta_w$), each occurring with equal probability, while player~2 has no private information. At the beginning of the game, nature draws player 1’s type, which is observed only by player~1.

The game unfolds over three stages. In stage 1, player 1 chooses to either threaten (\emph{T}) player 2, or do nothing and leave the situation at the status quo (\emph{SQ}). If \emph{SQ}, the game ends and each player receives a payoff of 0; if \emph{T}, then the game moves to stage 2 and player 2 chooses to either escalate (\emph{E}) the crisis or concede (\emph{C}). If \emph{C}, the game ends and the payoffs are $(1,-1)$; if \emph{E}, then the game moves to the final stage 3 and player 1 chooses to either attack (\emph{A}) or back down (\emph{BD}). If \emph{BD}, the game ends and the payoffs are $(-1,1)$; if \emph{A}, the game ends in a war, with payoffs $(2,-2)$ if player 1 is strong and $(-2,2)$ if player 1 is weak.



The (unique) sequential equilibrium of this game can be characterized by backward induction and Bayes' rule. 
In stage 3, sequential rationality requires the strong type of player 1 to choose \emph{A} and the weak type to choose \emph{BD}.
Anticipating this, the strong type chooses \emph{T} in stage 1, since doing so yields a positive continuation payoff. 
By contrast, the weak type mixes between \emph{T} and \emph{SQ} in equilibrium.\footnote{\label{fn:weak-mixing} If the weak type always chose \emph{T}, player 2 would choose \emph{E}, making the weak type prefer to deviate to \emph{SQ}. If the weak type always chose \emph{SQ}, player 2 would choose \emph{C} after observing \emph{T}, making the weak type prefer to deviate to \emph{T} for a positive payoff.}
The weak type is indifferent when player 2 mixes equally between \emph{C} and \emph{E}. 
Player 2 is indifferent when the posterior probability that player 1 is strong is $\frac{2}{3}$, which obtains by Bayes' rule when the weak type mixes equally between \emph{T} and \emph{SQ}.

As a first step in comparing CSE and SCE, we focus in this section on the fully cursed case, where $\chi = 1$ for CSE and $\phi = 1$ for SCE. We will revisit this game in Section~\ref{subsec:public_history} to analyze how general $\chi$-CSE and $\phi$-SCE differ. Before discussing the different notions of cursedness, we note that since sequential rationality is built into both CSE and SCE, the two solution concepts predict the same behavior in stage 3: strong player 1 chooses \emph{A}; weak player 1 chooses \emph{BD}.
Therefore, we will focus our analysis on behavior in the first two stages. In particular, we compare the notions of cursedness underlying the two solution concepts by addressing three questions: (i) Which player is cursed, in the sense of holding incorrect perceptions about the opponent’s strategy? (ii) How do these misperceptions distort beliefs about the opponent’s strategy? (iii) Why do the different forms of misperception under the two solution concepts lead to different predictions?

\begin{itemize}
    \item \textbf{Which player is cursed?} In this example, \ul{player~2 is the cursed player under both CSE and SCE}, which allows for a clean comparison between the two solution concepts.\footnote{In general, which player is cursed differs across the two solution concepts. For instance, in the signaling game example in Section~\ref{subsec:public_history}, the sender is cursed under SCE, whereas the receiver is cursed under CSE.} Under CSE, because player~1 is the only player with private information, cursedness applies to player~2 but not to player~1. Under SCE, cursedness also applies only to player~2, albeit for a different reason. Specifically, in stage~2, player~2 cannot distinguish among the possible histories leading to that stage and therefore cannot condition his action on them. As a result, cursedness under SCE does not affect player~1’s perception about player~2’s future stage 2 behavior at stage~1.

    \item \textbf{How is player 2 cursed?} \ul{Under SCE, player~2 mispredicts future actions but infers correctly from observed play; under CSE, both inference and prediction are distorted.}
    From the perspective of CSE, a fully cursed player completely neglects the dependence of actions on private types and thus never updates their belief about other players’ types. As a result, at stage 2, player 2 uses the prior to compute the average behavioral strategy at future stages, incorrectly believing that both types of player 1 will choose \emph{A} and \emph{BD} with equal probability in stage 3. 

    In contrast, under SCE, player~2 forms the correct Bayesian posterior belief about player~1’s private type but misperceives that player~1’s future actions do not depend on the history. Consequently, as in CSE, a fully cursed player~2 incorrectly believes that both types of player~1 will use the same average strategy in stage~3. However, unlike CSE, in SCE player~2 uses the \textit{correct} updated Bayesian posterior belief about player~1’s type to compute the average strategy of 
    player~1 
    in stage~3.

    \item \textbf{Why do different notions of cursedness generate different predictions?}
    \ul{CSE is a separating equilibrium, whereas SCE is a hybrid equilibrium},
    as summarized in Figure \ref{fig:crisis_bargaining_example}.
    Under CSE, player~2 incorrectly believes that both types of player~1 will choose \emph{A} and \emph{BD} with equal probability in stage~3, leading player~2 to choose \emph{E} in stage~2, since it yields an expected payoff of $0.5 > -1$, the payoff from choosing \emph{C}. Anticipating player~2’s choice of \emph{E}, weak
    player~1 will choose \emph{SQ} in stage~1, while strong player~1 will choose \emph{T}.
    
    However, this strategy profile cannot be supported as an SCE since, with SCE beliefs, player~2 would correctly infer that player~1 is surely type $\theta_s$ if \emph{T} were chosen. This would lead player 2 to
    choose \emph{C} in stage 2, since player~2 would expect player~1 to choose \emph{A} with certainty in stage~3. Instead, the only way to support the SCE is for weak player~1 to mix between \emph{T} and \emph{SQ} in stage~1, and player~2 must also mix between \emph{E} and \emph{C} to keep weak player~1 indifferent.
    Since player 1 is not cursed, the weak type's indifference condition is unchanged, so player 2 mixes between \emph{E} and \emph{C} with equal probability as in the sequential equilibrium. 
    However, under SCE, weak player 1 chooses \emph{T} less frequently than in the sequential equilibrium. 
    This raises player 2's posterior belief that player 1 is strong after observing \emph{T}, which offsets player 2's misperception that the weak type may choose \emph{A} and the strong type may choose \emph{BD} in stage 3.\footnote{The exact stage 1 SCE mixing probabilities are shown in Figure \ref{fig:crisis_bargaining_example} and derived in Proposition \ref{prop:crisis_bargaining_sce}.}  $\Diamond$

\end{itemize}

To summarize, this example illustrates how different notions of cursedness imply distinct equilibrium predictions, even though the cursed player is the same in both CSE and SCE.
Under CSE, the cursed player 2 fails to account for how player 1’s actions depend on their private type, leading to neglect of the information contained in observed actions and to an incorrect conjecture of player 1’s future behavior as an average strategy computed using the prior belief over types.
Under SCE, by contrast, the cursed player 2 correctly infers player 1’s type from past observed actions but fails to recognize how player 1’s unobserved, hypothetical future actions may depend on the history of play, namely, the realization of nature’s move in this example.
As a result, cursedness under SCE leads player 2 to misconjecture that player~1’s stage 3 action follows an average strategy, computed using the correct Bayesian posterior over player 1’s type.
In the next section, we will further examine how, more generally, $\chi$-CSE and $\phi$-SCE give rise to different belief updating dynamics, with formal illustrations.

\subsection{Differences in the Belief Updating Dynamics}
\label{subsec:belief_updating_new}

In addition to different notions of cursedness, another fundamental difference between the two solution concepts concerns the evolution of posterior beliefs about other players’ types, i.e., the belief-updating process regarding nature’s strategy. In the framework of CSE, players update their beliefs about the other players’ type profile \emph{at each public history} using the $\chi$-cursed Bayes’ rule. In the case of totally mixed strategies, the evolution of posterior beliefs under CSE can be characterized by Proposition \ref{lemma:rearrange}.

\begin{proposition}[Lemma 1 in \citealp{fong2023cursed}]\label{lemma:rearrange}
For any assessment $(\mu, \sigma)$ satisfying the $\chi$-cursed Bayes' rule, with $\sigma$ being a totally mixed behavioral strategy profile, for any $i \in N$, any non-terminal history $h^{t} = (h^{t-1}, a^{t}) \in \mathcal{H} \backslash \mathcal{H}^{T}$, and any $\theta \in \Theta$,
\begin{equation*}
\mu(\theta_{-i}| h^{t},  \theta_{i}) = \chi\mu(\theta_{-i}| h^{t-1}, \theta_{i}) + (1-\chi)\left[  \frac{\mu(\theta_{-i}| h^{t-1}, \theta_{i})\sigma_{-i}(a_{-i}^{t}|h^{t-1}, \theta_{-i})}{\sum_{\theta^{\prime
}_{-i}}\mu(\theta^{\prime}_{-i}|h^{t-1}, \theta_{i})\sigma_{-i}(a_{-i}%
^{t}|h^{t-1}, \theta^{\prime}_{-i})} \right]  .
\end{equation*}
\end{proposition}



In contrast, in the framework of SCE, the mathematical object corresponding to the posterior belief about other players’ types is the \emph{conjecture about nature’s strategy}, which is defined as a mixture of the cursed-plausible conjecture (with weight $\phi$) and the Bayesian conjecture (with weight $1-\phi$). Fix any totally mixed behavioral strategy profile $\sigma$ and any information set $I_i = (h^{t}, \theta_i)$, where $h^{t} = (h^{t-1}, a^{t})$. Under $\phi$-SCE, player $i$’s belief about nature’s initial move is
$$\mu^{\phi\mbox{-SCE}}(\theta_{-i}|
h^t, \theta_i) \equiv \phi 
\underbrace{\bar{\sigma}_\lambda^{I_i}(\theta | h_\emptyset)}_{\mbox{cursed}} +
(1-\phi)\underbrace{\sigma_\lambda^{I_i}(\theta|h_\emptyset)}_{\mbox{Bayesian}}.$$
As shown in Proposition \ref{lemma_sce_belief}, for any $\phi \in [0,1]$, $\phi$-SCE players always hold the correct Bayesian posterior, since the cursed-plausible conjecture coincides with the Bayesian conjecture.


\begin{proposition}\label{lemma_sce_belief} 
For any totally mixed behavioral strategy profile $\sigma$, any non-terminal information set $I_i=(h^{t}, \theta_{i})$, where $h^{t} = (a^{1}, \ldots, a^{t})$, 
any $\theta\in\Theta$, and for any $\phi \in [0,1]$, the cursed-plausible conjecture coincides with the Bayesian conjecture, and hence
$$\mu^{\phi\mbox{-SCE}}(\theta_{-i}|
h^t, \theta_i) = \frac{F\left(
\theta_{-i} | \theta_{i}\right)  \left[  \prod_{j \neq i} \prod_{1 \leq l \leq
t} \sigma_{j}\left(  a_{j}^{l} | h^{l-1}, \theta_{j}\right)  \right]  }%
{\sum_{\theta_{i}^{\prime} \in\Theta_{-i}} F\left(  \theta
_{-i}^{\prime} | \theta_{i}\right)  \left[  \prod_{j \neq i} \prod_{1 \leq l
\leq t} \sigma_{j}\left(  a_{j}^{l} |  h^{l-1}, \theta_{j}^{\prime}\right)
\right]  }.$$

\end{proposition}

The results from Propositions \ref{lemma:rearrange} and \ref{lemma_sce_belief} sharply highlight the differences in the belief-updating processes of the two solution concepts. Under the framework of $\phi$-SCE, players make correct inferences from observed events but fail to understand the implications of hypothetical events. Therefore, at any information set, since nature has already made its move, players always form the correct Bayesian posterior about other players’ types. In contrast, under the framework of $\chi$-CSE, players update their beliefs via Bayes’ rule while holding incorrect perceptions about other players’ behavioral strategies. That is, players do not make correct inferences from observed events. Consequently, as characterized by Proposition \ref{lemma:rearrange}, in $\chi$-CSE, players’ posterior beliefs about others’ types are dampened, being a weighted average of the posterior belief from the previous stage and the Bayesian posterior. In the following, we use a simple illustrative game to highlight this difference in belief-updating processes and to shed light on how they reflect distinct views of (cursed) learning behavior in dynamic games.

\bigskip
\noindent\textbf{Illustrative Example 2 (Broadcaster Game).}\footnote{This is adapted from a different
example proposed by Shengwu Li in private correspondence.}
Consider the following game with one broadcaster $(B)$ and two other players. There are two possible 
states---good $(\theta_g)$ or bad $(\theta_b)$---each occurring with equal probability. At the beginning of the game, nature draws the true state, which is observed only by the broadcaster and not by the other two players.
Therefore,
the true state is the broadcaster's private information.

In this game, all players except the broadcaster take turns choosing either \emph{In} or \emph{Out}. Each of these two players takes only one action, and all actions by all players are public. Moreover, before each player makes a move, the broadcaster makes a public announcement that the state is either good ($\theta_g$) or bad ($\theta_b$). In other words, this is a three-player, four-stage multistage game with observed actions: the broadcaster $B$ moves at the odd stages ($t = 1, 3$) with 
action set $A_B = \{G, B\}$, the first mover acts at stage two, and the second mover acts at stage four, with $A_1 = A_2 = \{In, Out\}$.

The broadcaster will get one unit payoff for each truthful announcement and 0
otherwise. Therefore, it is strictly optimal for the broadcaster to always
report the true state. For other players, they will get 0 for sure if they
choose \emph{Out}. If they choose \emph{In}, they will get
$\alpha\in(0,1)$ if the state is $\theta_{g}$ and $-1$ if the state is
$\theta_{b}$.

\begin{itemize}
\item\textbf{$\phi$-SCE:}  If players make correct inferences from observed events, they will recognize that the broadcaster’s action is \emph{perfectly correlated} with the true state. Accordingly, $\phi$-SCE predicts, for any $\phi \in [0,1]$, that both players will choose \emph{In} when the broadcaster announces $G$ and choose \emph{Out} otherwise.

\item\textbf{$\chi$-CSE:} In contrast, $\chi$-CSE posits that players fail to understand the dependence between private types and actions and therefore cannot recognize the perfect correlation between the broadcaster’s announcement and the true state. As a result, upon observing the broadcaster announce $G$, players do not update their belief to assign probability one to the true state being $\theta_g$. Instead, according to the $\chi$-cursed Bayes’ rule, their posterior belief that the true state is $\theta_g$ is 
$\mu^\chi(\theta_g | G) = 0.5\chi + (1-\chi) = 1 - 0.5\chi.$ Thus, even after observing the broadcaster announce $G$, the first mover will choose \emph{Out} when sufficiently cursed (i.e., when $\chi > \frac{2\alpha}{1 + \alpha}$).

Moreover, $\chi$-CSE players not only fail to understand the dependence between private types and actions, but also apply their incorrectly updated posterior beliefs when forming conjectures about other players’ future moves. In this example, after observing the broadcaster announce $G$, $\chi$-CSE players do not correctly predict that the broadcaster will announce $G$ again. Instead, they believe the broadcaster will announce $G$ again with probability $1 - 0.5\chi$, and expect the second mover to update their belief based on this mistaken perception. Consequently, $\chi$-CSE predicts that for $\chi \in \left[\frac{2\alpha}{1+\alpha}, \sqrt{\frac{2\alpha}{1+\alpha}}\right]$, the first mover will choose \emph{Out}, while the second mover will choose \emph{In}. 
\end{itemize}

To summarize, this example highlights the fundamental differences in 
belief-updating dynamics between CSE and SCE. Under $\phi$-SCE, players make correct inferences from observed events (for any $\phi \in [0,1]$), whereas $\chi$-CSE players fail to do so, with the parameter $\chi$ governing the degree of cursedness. Consequently, in SCE, collecting additional announcements from the broadcaster provides no new information for updating beliefs, as these announcements are perfectly correlated with the true state. In contrast, $\chi$-CSE players update their beliefs based on incorrect perceptions, causing them to learn the true state much more slowly. If players could observe infinitely many announcements, $\chi$-CSE predicts that (partially) cursed players would eventually learn the true state. $\Diamond$

\subsection{Differences in the Publicness of Public Histories}
\label{subsec:public_history}

Apart from their different notions of cursedness and belief updating dynamics, another significant difference
between CSE and SCE concerns the \emph{publicness} of public histories. In the CSE
framework, players correctly understand how choosing different actions leads to
different histories, and they know that all other players understand this as
well. In other words, from the perspective of CSE, \emph{public histories are
essentially public}.

In contrast, public histories are not necessarily public in the SCE framework.
The coarsest valid partition $\mathcal{P}$ requires only that histories in the same cell share the same active player and action set and respect perfect recall, so histories following \emph{different} public histories may be placed in the same coarsened information set. When this happens, a player believes that the other player acts in the same way regardless of which of these public histories has been reached. SCE players can therefore be cursed about endogenous public information, incorrectly believing that other players will not respond to changes in observed actions, including their own.

\begin{figure}[htbp!]
    \centering
    \includegraphics[width=0.8\linewidth]{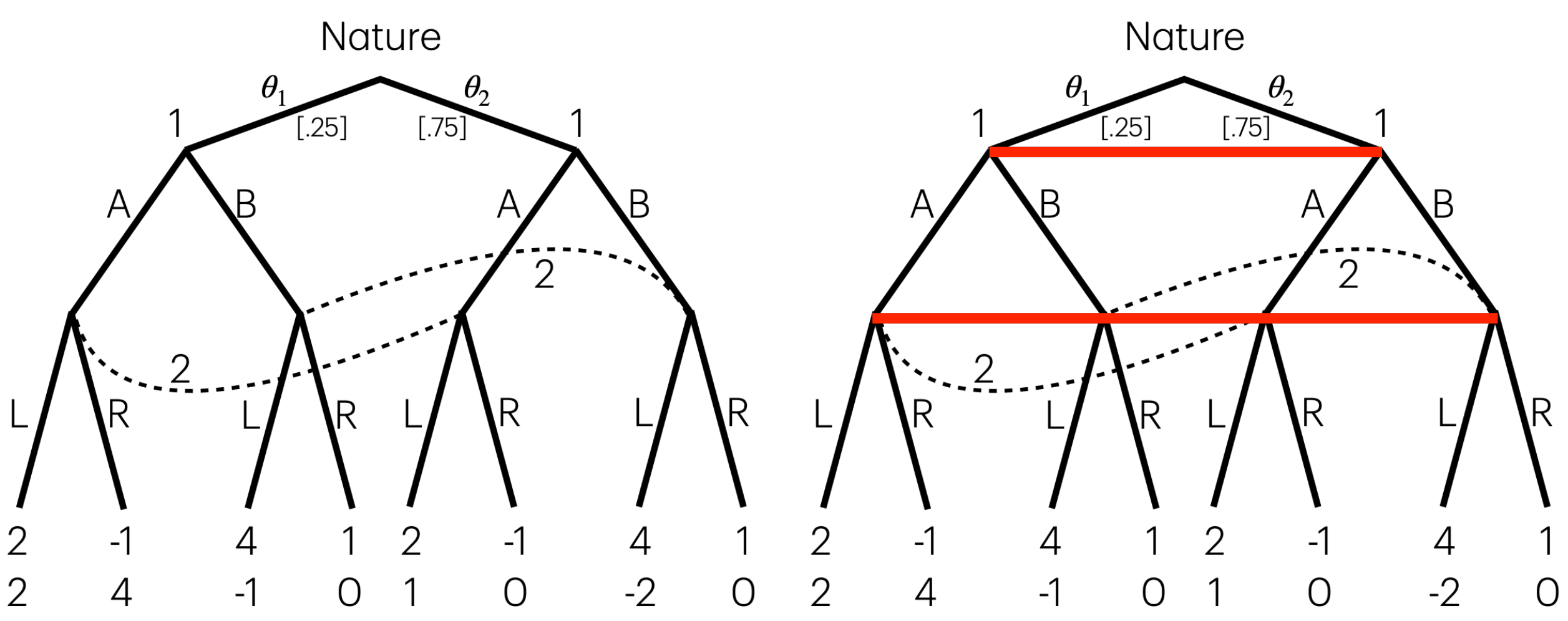}
    \caption{(Left) Game tree for the illustrative signaling game.
    (Right) The coarsest valid partition of the game (which violates PHC).}
    \label{fig:signaling_example}
\end{figure}

The signaling game in Figure \ref{fig:signaling_example},\footnote{This example is adapted from an earlier working paper version of “Cursed Sequential Equilibrium” \citep{fong2023cursed_working}. An analysis of a more general class of signaling games can be found in \cite{lin2025cursed}.} analyzed in detail in Section 
\ref{subsubsec:phc_violate}, illustrates how this can occur. 
In this game, the sender (player~1) has two possible types
drawn from the set $\Theta=\{\theta_{1},\theta_{2}\}$ with $F(\theta
_{1})=0.25$. The sender observes his type and sends a message  
$m\in\{A,B\}$, after which the receiver (player~2) observes the message and chooses an action $a\in\{L,R\}$. The receiver moves after two distinct public histories, $A$ and $B$. However, because she has the same action set $\{L,R\}$ after both messages and has not moved previously, perfect recall imposes no restriction on combining these histories. Consequently, the coarsest valid partition places all four of her histories, $\theta_1 A$, $\theta_1 B$, $\theta_2 A$, and $\theta_2 B$, in a single coarsened information set. An SCE cursed sender therefore perceives the receiver as responding identically to messages $A$ and $B$, even though the message he sends is publicly observed.

This example highlights the distinction between exogenous and endogenous information.
One motivation for SCE is to extend the notion of cursedness to capture players' failure to understand how other players' actions depend on the information sets they reach, which are \emph{endogenous} to the game. Under CSE, players neglect only the dependence between other players' actions and their \emph{exogenous} private information, while correctly understanding the dependence between other players' actions and endogenous public information. Under SCE, players may neglect both forms of dependence.
To gain deeper insights into the differences between the two solution concepts, it is instructive to distinguish between two classes of games based on whether the coarsest valid partition combines histories with different public histories. We refer to the condition that it does not do so as \emph{Public History Consistency (PHC)}.

\begin{definition}
[Public History Consistency (PHC)]\label{def_public} A multistage game with observed 
actions~$\Gamma$ satisfies Public History Consistency if its coarsest valid partition, $\mathcal{P}$, satisfies the following condition for all
$t < T,i\in N,\theta_{i}\in\Theta_{i},h^{t}\in\mathcal{H}^{t},\hat{h}^{t}\in\mathcal{H}^{t}$:
\[
h^{t}\neq\hat{h}^{t}\implies P(h^{t}, \theta_{i}) \cap
P(\hat{h}^{t}, \theta_{i}) = \emptyset.
\]
\end{definition}


In the signaling game above, PHC is violated because the receiver's coarsened information set contains histories with two distinct public histories, $A$ and $B$. By contrast, the 
crisis-bargaining game in Section \ref{subsec:cursedness_notion} satisfies PHC because each player moves after only one public history at each stage. Thus, there are no distinct public histories at the same stage that could be pooled into the same coarsened information set. 
In the following proposition, we provide a formal characterization of PHC.


\begin{proposition}\label{lemma:phc}
A multistage game with observed actions satisfies 
Public History Consistency if and only if its coarsest 
valid partition $\mathcal{P}$ satisfies that 
for any non-terminal information set $I_i = (h^t, \theta_i)$, 
$P(I_i) = \{(h^t, \theta'): \theta'\in \Theta \}.$

\end{proposition}

Proposition \ref{lemma:phc} provides a clear characterization of the class of multistage games with observed actions that satisfy PHC: PHC holds if and only if each cell of the coarsest valid partition contains exactly those histories that share the same public history and differ only in their type profiles.
In other words, in games that satisfy PHC, the coarsest valid partition is measurable with respect to public histories. Cursed players under both CSE and SCE understand that other players’ strategies can respond to their actions but fail to fully account for how other players’ private types may lead to different behaviors, albeit in different ways under the two solution concepts.

The remainder of this section examines how CSE and SCE differ depending on whether PHC holds. Section \ref{subsubsec:phc_violate} considers games in which PHC is violated. We show that SCE players’ failure to recognize the publicness of public histories leads to predictions that differ sharply from those of CSE, including in games of complete information. Section \ref{subsubsec:phc_hold} then examines games in which PHC holds and shows that CSE and SCE can still generate different predictions because the two concepts imply different inferences from observed events.

\subsubsection{When Public History Consistency Is Violated}
\label{subsubsec:phc_violate}

PHC may seem like an innocuous property, especially in multistage games with observed actions, where all actions are perfectly monitored.
However, PHC is violated in many games, such as the signaling game depicted in Figure~\ref{fig:signaling_example}.\footnote{Alternatively, one could modify the definition of SCE by \emph{requiring} the underlying partition to be consistent with public histories, even in games that do not satisfy PHC. This gives rise to a different equilibrium concept, which could be called \emph{Public Sequential Cursed Equilibrium (PSCE)}.} 
To demonstrate how the different treatments of public histories under CSE and SCE can lead to different predictions when PHC is violated, we revisit this signaling game.

\bigskip
\noindent\textbf{Illustrative Example 3 (Signaling Game).} For simplicity, we focus on pure-strategy equilibria and use the four-tuple $[(m(\theta_{1}), m(\theta_{2})); (a(A), a(B))]$ to denote a behavioral strategy profile. Table \ref{tab:solution_comp} summarizes the sequential equilibrium, standard cursed equilibrium, $\chi$-CSE, and $\phi$-SCE of the game. The derivations of $\chi$-CSE and $\phi$-SCE are provided in \ref{appendix:proofs}.

\begin{table}[htbp!]
\centering
\caption{Predictions of Different Solution Concepts of the Signaling Game}
\label{tab:solution_comp}
\renewcommand{\arraystretch}{1.8}
\begin{tabular}{cccccc}
\hline
 &  & \multicolumn{4}{c}{Solution Concepts} \\ \cline{3-6} 
Strategy Profile &  & SE & $\chi'$-CE & $\chi$-CSE & $\phi$-SCE \\ \hline
$[(A,A); (L,R)]$ &  & Yes & $\chi'\in[0,1]$ & $\chi\in[0,1]$ & $\phi\leq 1/3$ \\
$[(B,B); (R,R)]$ &  & Yes & $\chi'\in[0,1]$ & $\chi\leq 8/9$ & $\phi\in[0,1]$ \\
$[(B,B); (L,R)]$ &  & No & No & No & $\phi\geq 1/3$ \\
$[(B,A); (L,R)]$ &  & No & No & No & $\phi = 1/3$ \\ \hline
\end{tabular}
\end{table}

To analyze this game, we first note that upon observing $B$, it is strictly dominant for player~2 to choose $R$. Moreover, since the incentives of the two types of player~1 are identical, there is no separating sequential equilibrium. The only pure sequential equilibria are pooling equilibria, in which the two types of player~1 pool at $A$ and at $B$, respectively. Lastly, by Proposition~3 of \cite{eyster2005cursed}, these two equilibria are in fact pooling $\chi'$-CE for any $\chi' \in [0,1]$.

When extending the framework of CE to dynamic games, the publicness of public histories plays an important role, as is clearly illustrated by this game. Under $\chi$-CSE, cursed player~1 knows that player~2’s strategy is contingent on the message he sends; that is, player~1 understands that player~2 may behave differently depending on whether he chooses $A$ or $B$. Moreover, under $\chi$-CSE, player~1 also knows that player~2 will draw incorrect inferences using the $\chi$-cursed Bayes’ rule, which causes the equilibrium $[(B,B); (R,R)]$ to break down when $\chi$ is sufficiently large. We refer readers to the working paper version of ``Cursed Sequential Equilibrium'' \citep{fong2023cursed_working} for a detailed discussion.

In contrast, under the framework of $\phi$-SCE, cursed player~1 incorrectly believes that player~2 will choose the same action at the off-path information set as at the on-path information set, as player~1 fails to recognize the publicness of his action. As a result, when $\phi$ is sufficiently large, there exists an equilibrium $[(B,B); (L,R)]$ that is neither a sequential equilibrium, nor a $\chi'$-CE, nor a $\chi$-CSE. This strategy profile is supported as a $\phi$-SCE by cursed player~1’s incorrect perception that player~2 will also choose $R$ at the off-path information set $A$.
Lastly, it is worth remarking that there exists a knife-edge case in which $[(B,A); (L,R)]$ is a separating $\phi$-SCE if and only if $\phi = 1/3$. Since the incentives of both types of player~1 are perfectly aligned, the existence of such a separating $\phi$-SCE is unexpected. $\Diamond$

\bigskip



The implications of PHC violations are not limited to games with private information.
In particular, CSE and SCE can generate different predictions in games of complete information, that is, when $|\Theta| = 1$.
In the 
CSE framework, players fail only to understand the dependence between 
private types and actions. Consequently, in games of complete information,
CSE coincides with sequential equilibrium, as shown in 
Proposition~\ref{claim:ci_cse} below. 
Because the type space is a singleton, there is no possibility
for players to make mistaken inferences about the types of other players, 
rendering the neglect of the correlation between types and actions moot, as
is also the case for CE.

\begin{proposition}\label{claim:ci_cse}
If $|\Theta| = 1$, $\chi$-CSE is equivalent to
sequential equilibrium for any $\chi\in[0,1]$.    
\end{proposition}



In contrast, SCE does not necessarily coincide with sequential equilibrium even in games of \textit{perfect information}, where nature is not a player and every information set is a singleton. 
The following example illustrates how PHC can be violated even in such games: the coarsest valid partition coarsens distinct public histories into the same information set, leading SCE players to misunderstand how their current actions affect other players' future behavior.

\bigskip
\noindent\textbf{Illustrative Example 4 (Perfect-Information Game).}\footnote{This example is adapted from the
game depicted in Figure 6 of \cite{cohen2022sequential}.}
Consider the two-player game of perfect information in Figure
\ref{fig:tree_complete}. In the first stage, player 1 makes a choice from
$A_{1}=\{B,R\}$. After observing player 1's decision, player 2 then makes a
choice from $A_{2}\in\{b,r\}$. The payoffs are shown in the game tree where
$x\in\mathbb{R}$ and $y<1$. In the following, we will denote the players'
(pure) behavioral strategy profile by $[a_{1};(a_{2}(B),a_{2}(R))]$.

\begin{figure}[ptbh]
\centering
\par
\tikzset{
solid node/.style={circle,draw,inner sep=1.5,fill=black},
hollow node/.style={circle,draw,inner sep=1.5}
} \begin{tikzpicture}[scale=1.5,font=\footnotesize]
\tikzstyle{level 1}=[level distance=15mm,sibling distance=35mm]
\tikzstyle{level 2}=[level distance=15mm,sibling distance=15mm]
\node(0)[solid node,label=above:{$1$}]{}
child{node(1)[solid node,label=above:{$2$}]{}
child{node[label=below:{$(2,2)$}]{} edge from parent node[left]{$b$}}
child{node[label=below:{$(0,0)$}]{} edge from parent node[right]{$r$}}
edge from parent node[left,xshift=-3]{$B$}
}
child{node(2)[solid node,label=above:{$2$}]{}
child{node[label=below:{$(x,y)$}]{} edge from parent node[left]{$b$}}
child{node[label=below:{$(1,1)$}]{} edge from parent node[right]{$r$}}
edge from parent node[right,xshift=3]{$R$}
};
\end{tikzpicture}
\caption{A Two-Player Game with Complete Information}%
\label{fig:tree_complete}%
\end{figure}

The unique subgame-perfect Nash equilibrium of this game is
$[B; (b, r)]$, yet $[R; (b, r)]$ can also be supported as a 
$\phi$-SCE if $\phi \geq 1/2$. To verify this $\phi$-SCE, we first 
note that the coarsest partition of the game satisfies
$P(h^1 = B) = P(h^1 = R) = \{B, R\}$. Given $[R; (b, r)]$, player 1 conjectures
that player 2 when observing $B$ will choose $r$ with probability 
$\phi$ and choose $b$ with probability $1-\phi$.
Therefore, player 1 will not have an incentive to deviate to $B$ if and only if
$1 \geq 2(1-\phi) \iff \phi \geq 1/2.$

It is worth noting that, for any $x\in\mathbb{R}$ and $y<1$, $R$ can be supported as a $\phi$-SCE strategy for player 1 whenever $\phi\geq 1/2$. 
Thus, this additional SCE equilibrium persists across the entire range of admissible values of $x$ and $y$. $\Diamond$

\subsubsection{When Public History Consistency Is Satisfied}
\label{subsubsec:phc_hold}

The difference in games of complete information is 
due to the coarsening of public histories into the same information sets in SCE. 
The requirement of an information set is
that the set of actions for every history in this information set is exactly
the same. Therefore, if the actions at every public history are all labeled
differently, then the coarsest valid partition is consistent with the public
histories. This observation is proved in Proposition \ref{claim:label_claim}.

\begin{definition}
A multistage game with observed actions is \textbf{scrambled} if for any
$i\in N$, any $t<T$, and any $h, h^{\prime}\in\mathcal{H}^{t}$ such that
$h\neq h^{\prime}$, then
\[
\forall\; s\in A_{i}(h) \;\mbox{ and }\; s^{\prime}\in A_{i}(h^{\prime})
\implies s\neq s^{\prime}.
\]

\end{definition}

\begin{proposition}\label{claim:label_claim} 
A scrambled multistage game with observed actions
satisfies PHC.
\end{proposition}

Proposition \ref{claim:label_claim} implies that whether a game satisfies PHC depends in part on how its actions are labeled, suggesting a simple way to apply SCE when label-invariant predictions are desired. In SCE, action labels are part of the model's primitives and determine which histories can be pooled into the same coarsened information set (see \cite{cohen2022sequential} for a discussion of this modeling choice). Researchers who prefer predictions that do not depend on these labels can first scramble the game and then apply SCE. By 
Proposition~\ref{claim:label_claim}, the scrambled game satisfies PHC.\footnote{\cite{cohen2022sequential} also propose \emph{causal SCE} (see their Appendix F), under which each player correctly understands how their own actions affect other players’ subsequent behavior. This provides another way to address relabeling across different public histories.} Under CSE, this invariance holds automatically because the average behavioral strategy is defined at each public history. This difference reflects distinct modeling choices rather than a limitation of either concept.
To illustrate how scrambling affects SCE predictions in practice, we return to the signaling game from Section \ref{subsubsec:phc_violate}.

\begin{figure}[htbp!]
    \centering
    \includegraphics[width=0.8\linewidth]{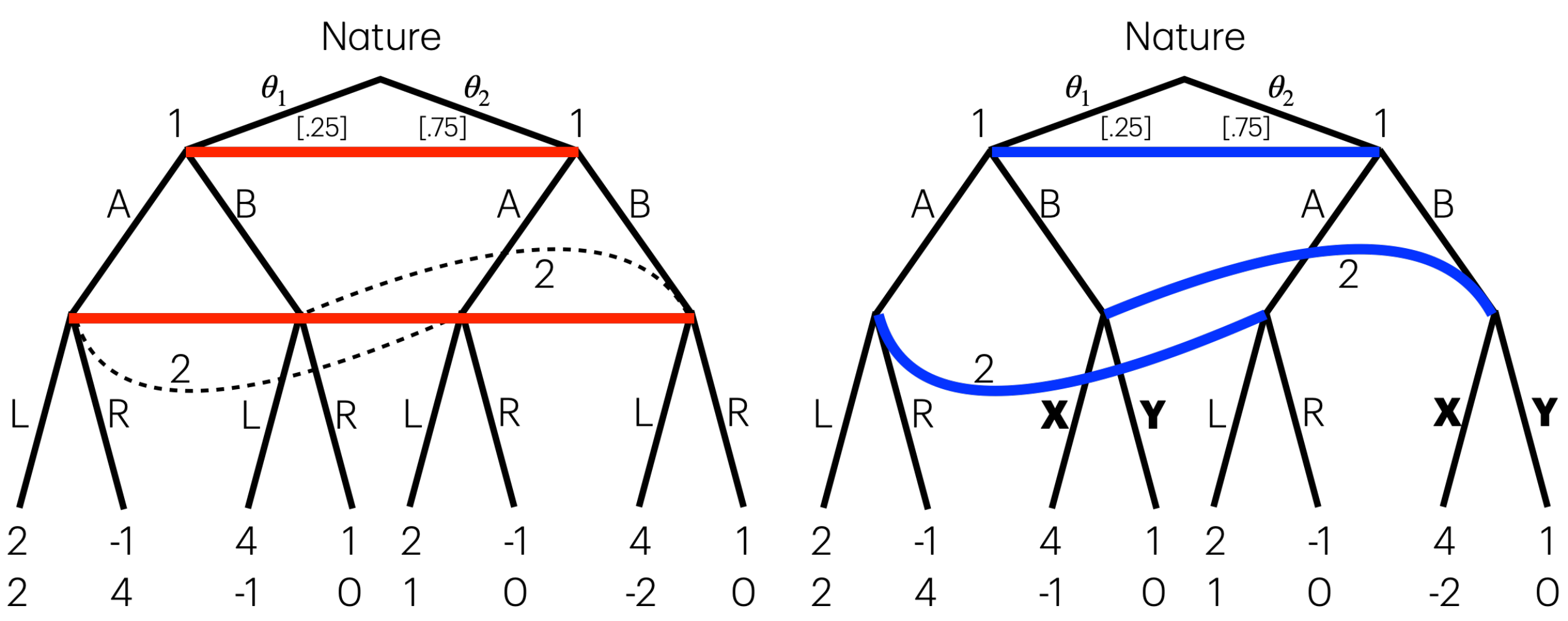}
    \caption{(Left) Game tree for the original signaling game and its coarsest valid partition.
    (Right) The coarsest valid partition of the re-labeled game (which satisfies PHC).} 
    \label{fig:signaling_example_relabel}
\end{figure}

\bigskip
\noindent\textbf{Illustrative Example 5 (Re-labeled Signaling Game).}
The earlier example of signaling game can also demonstrate the effect of re-labeling. Let $A_{2}(A)$ and $A_{2}(B)$ denote the action sets at the public histories $A$ and $B$, respectively. If $A_{2}(A) = \{L, R\}$ and $A_{2}(B) = \{X, Y\}$, then the game is scrambled and therefore satisfies PHC. As a result, the coarsest valid partition of the re-labeled signaling game coincides with the partition shown in the right panel of Figure \ref{fig:signaling_example_relabel}.

In contrast to the original game, which admits four pure-strategy $\phi$-SCE---including two equilibria that are not sequential equilibria---the re-labeled game yields the same predictions under $\phi$-SCE as under sequential equilibrium. In particular, $[(A, A); (L, Y)]$ and $[(B, B); (R, Y)]$ are the only $\phi$-SCE for any $\phi \in [0,1]$ in the 
re-labeled game. 

The intuition behind this result is that, once the actions are re-labeled, the coarsest valid partition becomes consistent with public histories, which effectively eliminates player 1’s misperception that player 2 will take the same action across different public histories. In this example, \emph{regardless of which message he actually sends}, player 1 correctly anticipates that player 2 would choose $Y$, the dominant action, following message $B$. As a result, an SCE player 1 in the 
re-labeled game engages in the same reasoning as in sequential equilibrium, correctly understanding that player 2’s strategy is contingent on public histories. Consequently, the strategy profiles $[(B, B); (L, R)]$ and $[(B, A); (L, R)]$, which were supported as SCE in the original game due to player 1’s misunderstanding that player 2’s strategy does not condition on public histories, can no longer be supported in the re-labeled game. $\Diamond$

\bigskip

When a game satisfies PHC, players under both solution concepts correctly understand that other players' strategies can depend on public histories. Nevertheless, PHC does not eliminate all differences between CSE and SCE.
This difference persists because cursed players draw different inferences from observed events and use different posterior beliefs when forming conjectures about other players’ hypothetical actions. Under CSE, cursed players hold distorted posterior beliefs about other players’ types and apply these \emph{distorted} beliefs to compute the average behavioral strategy of other players. In contrast, under SCE with PHC, cursed players make correct Bayesian inferences from observed events. Consequently, when forming their conjectures, they use the correct \emph{Bayesian} posterior to compute the average behavioral strategy for every other player. To illustrate how these differences lead to different predictions, we revisit the leading example from Section \ref{subsec:cursedness_notion}, in which PHC is satisfied.

Before diving into the illustrative example, Remark \ref{remark_past_future} derives a player’s cursed-plausible conjectures about other players’ strategies in past and future events under SCE with PHC. This remark will be useful for the calculations in the illustrative example.

\begin{remark}
\label{remark_past_future}

Fix any non-terminal information set $I_i = (h^{t}, \theta_{i})$ such that there exist public histories $h^{t^{\prime}}$ and $h^{t^{\prime\prime}}$ satisfying $h^{t^{\prime}} \prec h^{t} \preceq h^{t^{\prime\prime}}$. Let
$h^{t^{\prime\prime}} \equiv (a^{1}, \ldots, a^{t^{\prime}}, \ldots, a^{t}, \ldots, a^{t^{\prime\prime}})$.
Under PHC, player $i$’s SCE cursed-plausible conjecture about type $\theta_{j}$ player $j$’s partial strategy at the past information set $I'_j = (h^{t^{\prime}}, \theta_j)$ is:
\begin{align*}
\bar{\sigma}_{j}^{I_{i}}(\tilde{a}_{j}^{t^{\prime}+1} |
I_j') =
\begin{cases}
1 & \text{ if } \tilde{a}_{j}^{t^{\prime}+1}=a_{j}^{t^{\prime}+1}\\
0 & \text{ if } \tilde{a}_{j}^{t^{\prime}+1} \neq a_{j}^{t^{\prime}+1}.%
\end{cases}
\end{align*}
On the other hand, player $i$'s cursed-plausible
conjectures about type $\theta_{j}$ player
$j$'s partial strategy at the future information set 
$I''_j = (h^{t^{\prime\prime}}, \theta_j)$ is:
\begin{align*}
\bar{\sigma}_{j}^{I_{i}}(a_{j}^{t^{\prime\prime}+1}|
I_j'') = \sum_{\theta
^{\prime}_{j} \in\Theta_{j}} \mu^{*}(\theta^{\prime}_{j} | 
I_i) \sigma_{j}(a_{j}^{t^{\prime\prime}+1} |  I_j'')
\end{align*}
where $\mu^*$ is the correct Bayesian posterior (induced by 
strategy profile $\sigma$).\footnote{Note that $\mu^*(\theta'_j|I_i)$ is the Bayesian posterior conditional on $I_i$, the information set at which player $i$ is currently located. Since under SCE player $i$ fails to draw inferences from hypothetical events, her cursed-plausible conjectures about player $j$’s future partial strategy reflect a posterior belief that remains stuck at $I_i$. As player $i$ moves on, her conjecture, together with his posterior belief, is updated, which may lead player $i$ to deviate from her original contingent plan at $I_i$. For a detailed illustration of such \textit{dynamically inconsistent} behavior under SCE, see Example 2.8 and Appendix B of \cite{cohen2022sequential}.}
\end{remark}

\bigskip

\noindent\textbf{Illustrative Example 6 (Crisis-Bargaining Game Revisited).} To illustrate how $\chi$-CSE and $\phi$-SCE can differ even when PHC is satisfied, we revisit the crisis-bargaining game from Section \ref{subsec:cursedness_notion} and solve for its $\chi$-CSE and $\phi$-SCE. Since it is dominant for type $\theta_s$ player~1 to choose \emph{T} in stage 1 and \emph{A} in stage 3, and for type $\theta_w$ player~1 to choose \emph{BD} in stage 3, it suffices to characterize the equilibrium by solving for the probabilities with which type $\theta_w$ player~1 chooses \emph{T} in stage~1 and player~2 chooses \emph{E} in stage~2, denoted by $\sigma_{1W}$ and $\sigma_2$, respectively.


\begin{figure}[htbp!]
    \centering
    \includegraphics[width=\linewidth]{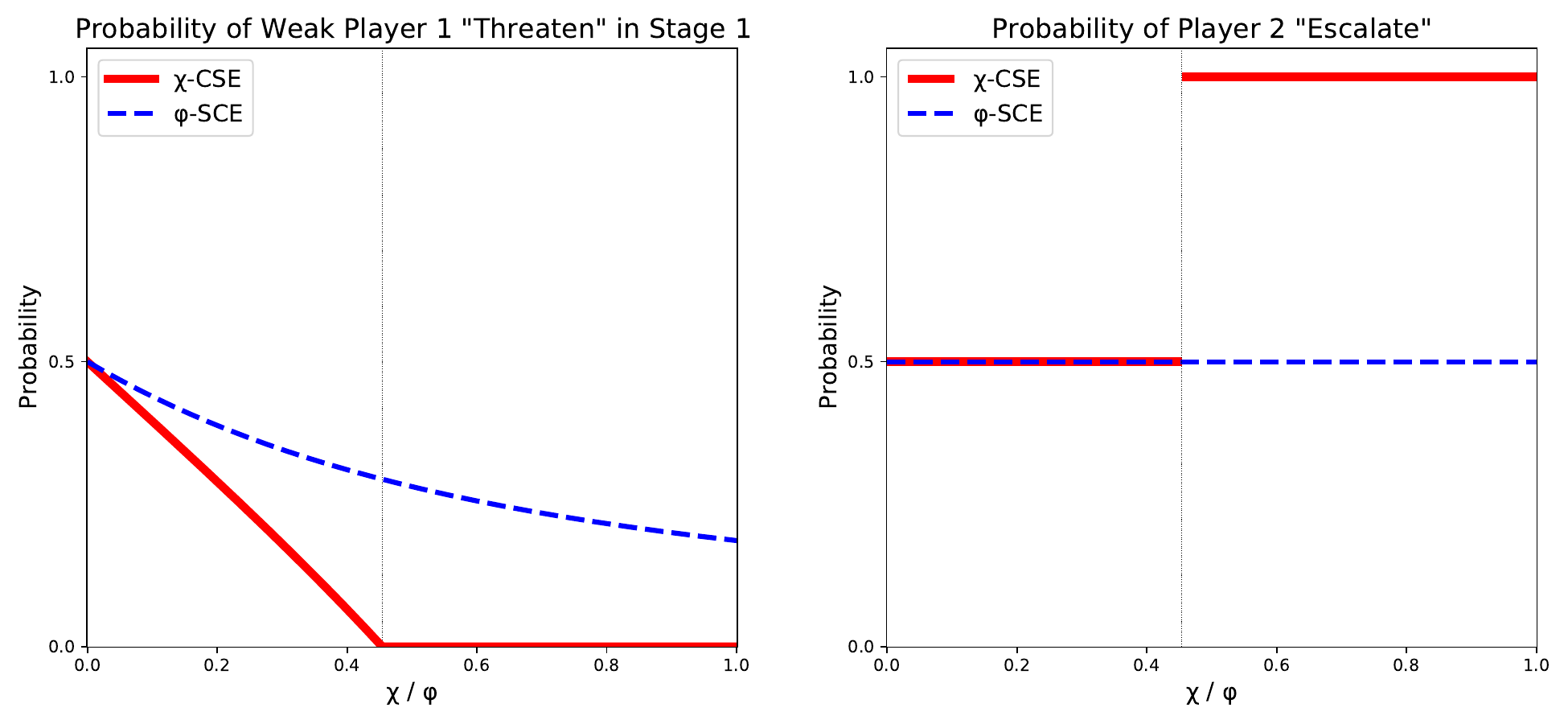}
    \caption{$\chi$-CSE and $\phi$-SCE equilibrium probabilities of type~$\theta_w$ player~1 choosing \emph{T} in stage~1 and player~2 choosing \emph{E} in stage~2.}
    \label{fig:crisis_eq}
\end{figure}

Figure~\ref{fig:crisis_eq} summarizes the predictions of $\chi$-CSE (solid lines) and $\phi$-SCE (dashed lines) for $\sigma_{1W}$ (left panel) and $\sigma_2$ (right panel) across different levels of cursedness.
The derivations of $\chi$-CSE and $\phi$-SCE are provided in \ref{appendix:proofs}.
Under $\chi$-CSE, there exists a critical value $\chi \approx 0.454$ at which the equilibrium structure changes.
For $\chi$ below this threshold, $\chi$-CSE is a hybrid equilibrium, in which the probability that type $\theta_w$ player~1 chooses \emph{T} in stage~1 decreases from $0.5$ to $0$ as $\chi$ approaches the critical value.
For $\chi$ above this threshold, $\chi$-CSE is a pure separating equilibrium, coinciding with the fully cursed case.
In contrast, $\phi$-SCE is a hybrid equilibrium for all $\phi \in [0,1]$, with the probability that type $\theta_w$ player~1 chooses \emph{T} in stage~1 decreasing smoothly from $0.5$ to approximately $0.186$ as $\phi$ increases to $1$.

To clarify the intuition behind these predictions, we begin by focusing on the sequential equilibrium of the game, corresponding to $\chi = 0$ or $\phi = 0$, as a benchmark.
Note that type $\theta_w$ player~1 and player~2 are effectively playing a matching-pennies game (see footnote~\ref{fn:weak-mixing}), in which player~2 aims to match while player~1 aims to mismatch, resulting in a mixed-strategy equilibrium.
In this equilibrium, $\sigma_{1W}=\sigma_2=0.5$, making each other indifferent between the two available actions.

As $\chi$ or $\phi$ increases, player~2 in stage~2 becomes partially cursed and incorrectly conjectures that player~1’s contingent strategy in stage~3 shifts weight away from the actual, history-dependent strategy toward an average, history-independent strategy, rather than placing full weight on the former.
In other words, cursedness leads player~2 to incorrectly conjecture that type $\theta_s$ player~1 would choose \emph{BD} with positive probability and that type $\theta_w$ player~1 would choose \emph{A} with positive probability in stage 3, thereby inflating player~2’s expected payoff from choosing \emph{E} in stage 2. In turn, type $\theta_w$ player~1 reduces the probability of choosing \emph{T} to restore player~2’s indifference in equilibrium. This mechanism operates under both $\chi$-CSE and $\phi$-SCE, so $\sigma_{1W}$ decreases with respect to both $\chi$ and $\phi$, but at different rates. However, while the \textit{qualitative} effect of the mechanism is the same for $\chi$-CSE and $\phi$-SCE, both the logical reasoning for the effects and their quantitative magnitudes are different.

Under $\chi$-CSE, $\chi$-cursed player~2 also mistakenly believes that player~1’s past strategy in stage~1 places weight $\chi$ on the average behavioral strategy.
Consequently, even if type $\theta_w$ player~1 always chooses \emph{SQ}, player~2 believes that an observed \emph{T} at stage~2 may have been chosen by type $\theta_w$.
As illustrated in Section~\ref{subsec:cursedness_notion} (for the case $\chi=1$), this belief makes choosing \emph{E} profitable for player~2 when $\chi$ is sufficiently large and hence supports $\sigma_{1W} = 0$ in $\chi$-CSE.
Intuitively, cursedness under $\chi$-CSE distorts player~2’s perceptions along two dimensions in the matching-pennies-like interaction with type $\theta_w$ player~1: it alters perceptions of the payoff structure through distorted conjectures about future actions, as under $\phi$-SCE, and it also distorts beliefs about the opponent’s behavior through misperceptions of past actions.
This difference in the notions of cursedness explains why the discontinuity shown in Figure~\ref{fig:crisis_eq} arises under $\chi$-CSE but not under $\phi$-SCE, highlighting a sharp distinction between the two equilibrium concepts even in a game that satisfies PHC. $\Diamond$

\subsection{Differences in One-Stage Simultaneous-Move Games}

\label{subsec:one_stage}

The last important difference between CSE and SCE lies in their 
relationship with the standard cursed equilibrium in one-stage games. 
As shown by \cite{fong2023cursed}, CSE coincides with the standard CE for any 
one-stage game. In contrast, \cite{cohen2022sequential} show that SCE in 
one-stage games is equivalent to \emph{independently cursed equilibrium} (ICE), under 
which players are cursed not only about the dependence of opponents’ actions 
on private information but also about the correlation between opponents’ actions 
induced by correlation in the prior distribution. In what follows, we 
first summarize the definitions of CE and ICE and then illustrate, through 
an example of a three-player game, how substantially their predictions can differ.

Definition \ref{def:CE} describes the standard CE in an one-stage game. Under
CE, a player fails to account for how the other players' action profile may
depend on their types, and best responds to the \textit{average strategy
profile of all other players}.

\begin{definition}
[Cursed Equilibrium in \citealp{eyster2005cursed}]\label{def:CE} A strategy
profile $\sigma$ is a cursed equilibrium (CE) if for each player $i$, type
$\theta_{i}\in\Theta_{i}$ and each $a_{i}^{1}\in A_{i}(h_{\emptyset})$ such
that $\sigma_{i}(a_{i}^{1} | \theta_{i},h_{\emptyset})>0$,
\begin{align*}
a_{i}^{1} \in\argmax_{a_{i}^{1^{\prime}}\in A_{i}(h_{\emptyset})}  &
\sum_{\theta_{-i}\in\Theta_{-i}}F(\theta_{-i}|\theta_{i}) \;
\times\\
&  \qquad\qquad\sum_{a_{-i}^{1}\in A_{-i}(h_{\emptyset})} \left[  \sum
_{\theta_{-i}\in\Theta_{-i}}F(\theta_{-i}|\theta_{i})\sigma
_{-i}(a_{-i}^{1}|h_{\emptyset}, \theta_{-i})\right]  u_{i}(a_{-i}^{1}, a_{i}^{1^{\prime}}, 
\theta_{-i}, \theta_{i}).
\end{align*}

\end{definition}

Definition \ref{def:SCE} provides the definition of ICE in an one-stage game.
Under ICE, a player fails to account for how each player's action may depend
on her own type, and how it may correlate with another player's action (via
the correlation in type distribution). Therefore, a player would best respond
as if the \textit{average strategies across the other players are independent}.

\begin{definition}
[Independently Cursed Equilibrium in \citealp{cohen2022sequential}]\label{def:SCE} 
A strategy profile $\sigma$
is an independently cursed equilibrium (ICE) if for each player $i$, type
$\theta_{i}\in\Theta_{i}$ and each $a_{i}^{1}\in A_{i}(h_{\emptyset})$ such
that $\sigma_{i}(a_{i}^{1} | \theta_{i},h_{\emptyset})>0$,
\begin{align*}
a_{i}^{1} \in\argmax_{a_{i}^{1^{\prime}}\in A_{i}(h_{\emptyset})}  &
\sum_{\theta_{-i}\in\Theta_{-i}}F(\theta_{-i}|\theta_{i}) \;
\times\\
&  \qquad\quad\sum_{a_{-i}^{1}\in A_{-i}(h_{\emptyset})} \prod_{j \in
N\backslash\{i\}} \left[  \sum_{\theta_{j}\in\Theta_{j}}F(\theta
_{j}|\theta_{i})\sigma_{j}(a_{j}^{1}|h_{\emptyset}, \theta_{j})\right]
u_{i}(a_{-i}^{1}, a_{i}^{1^{\prime}}, \theta_{-i}, \theta_{i}).
\end{align*}

\end{definition}

Propositions \ref{prop:one_period} and \ref{prop:one_period_sce} summarize the
relations between CSE, SCE, and CE in one-stage games. The proofs can be found
in \cite{fong2023cursed} and \cite{cohen2022sequential}, respectively.

\begin{proposition}[Proposition 4 in \citealp{fong2023cursed}]\label{prop:one_period}
For any one-stage game and for any $\chi\in[0,1] $, $\chi$-CSE and $\chi$-CE are equivalent.   
\end{proposition}

\begin{proposition}[Theorem 2.9 in \citealp{cohen2022sequential}]\label{prop:one_period_sce} 
For any one-stage game, a strategy profile is an ICE if and only if there exists a
system of conjectures such that the resulting assessment is an SCE.
\end{proposition}

Although CE and ICE are equivalent in an one-stage two-person game, we show in
the following example that the sets of CE and ICE may be non-overlapping in a
three-person game. This finding suggests that CSE and SCE are generally
different when there are more than two players.

\bigskip

\noindent\textbf{Illustrative Example 7 (One-Stage Game).} Consider the following three-player
one-stage game. Player 1 and 2 have two
possible types drawn from the set $\Theta= \{b, r\}$ with the joint
distribution $F(\theta_{1} = \theta_{2} = b) = F
(\theta_{1} = \theta_{2} = r) = 0.4$ and $F(\theta_{1} =
b, \theta_{2} = r) = F(\theta_{1} = r, \theta_{2} = b) = 0.1$. 
Player 3 has no private information. Each player
makes a choice from the set $A = \{b, r, m\}$. Player 1 and 2 will get one
unit of payoff if his choice matches his type (and 0 otherwise). Player 3 will
get one unit of payoff if his choice matches player 1's and 2's choices when
$a_{1} = a_{2}$, or if he chooses $m$ when $a_{1} \neq a_{2}$ (and 0 otherwise).

To summarize, player 1 and 2 have private information, and their payoffs will
be maximized if their actions match their types. Player 1's (and 2's) type is
$b$ or $r$ with equal probabilities. However, their types can be the same with
probability 0.8. Player 3 has no private information, and his goal
is to guess his opponents' actions by following them when they act the same
and choosing $m$ when they act differently.

In this game, CE differs from ICE in that player 3 will choose $b$ or $r$ under CE,
whereas he will choose $m$ under ICE. Because choosing the same action as 
their private type is a strictly dominant strategy for player 1 and 2, under CE,
player 3’s expected payoff from choosing $a_3 = b$ or $r$ is $F(\theta_1 = \theta_2 = a_3) = 0.4$, 
while the expected payoff from choosing $a_3 = m$ is $F(\theta_1 \neq \theta_2) = 0.2$. As a 
result, under CE, player 3 will choose either $b$ or $r$.

In contrast, under ICE, player 3’s expected payoff from choosing 
$a_3 = b$ or $r$ is $F(\theta_1 = a_3)\times F(\theta_2 = a_3) = 0.25$, 
whereas the expected payoff from choosing $a_3 = m$ is
$F(\theta_1 = b)\times F(\theta_2 = r) + F(\theta_1 = r)\times F(\theta_2 = b) = 0.5$.
Thus, under ICE, player 3 will choose $m$ rather than $b$ or $r$.
$\Diamond$

\section{Concluding Remarks}
\label{sec:conclusion}

This paper provides a systematic comparison of two recent extensions of Cursed Equilibrium (CE) to dynamic games: Cursed Sequential Equilibrium (CSE) by \cite{fong2023cursed} and Sequential Cursed Equilibrium (SCE) by \cite{cohen2022sequential}. While both concepts build on the seminal work of \cite{eyster2005cursed}, these extensions capture fundamentally different concepts of cursedness in dynamic strategic environments.

Under CSE, cursedness captures players’ failure to account for the correlation between opponents’ private types and actions, leading to distorted belief updating along the path of play. As a result, players draw incorrect inferences from observed events and form expectations about other players' future behavior based on these distorted posterior beliefs.
In contrast, SCE models cursedness as a failure to recognize how opponents’ unobserved actions depend on the history of play, while still maintaining correct inference from past observed events.
In multistage games with observed actions, cursed players under SCE update their beliefs correctly using Bayes’ rule, yet still make incorrect predictions about other players’ future moves.

The main contribution of this paper is to clarify the conceptual and theoretical differences between CSE and SCE. In addition to their distinct notions of cursedness, we identify several differences embedded in these approaches, including differences in belief-updating dynamics, the treatment of public histories, and equilibrium predictions in one-stage simultaneous-move games.
In particular, our analysis highlights how SCE's treatment of public histories can generate departures from sequential equilibrium even under complete information and make predictions sensitive to the re-labeling of actions.
We also illustrate how these distinct notions of cursedness can lead to substantially different predictions through examples involving signaling and crisis-bargaining games, thereby helping researchers understand and apply the two concepts in dynamic games.


\begin{singlespace}
\bibliographystyle{ecta}
\bibliography{reference}
\end{singlespace}

\newpage
\appendix

\renewcommand{\thesection}{Appendix}

\section{Omitted Proofs and Derivations}
\label{appendix:proofs}

\setcounter{proposition}{0}
\renewcommand{\theproposition}{A.\arabic{proposition}}

\subsection*{Proof of Proposition \ref{lemma_sce_belief}}

At $I_i = (h^t, \theta_i)$, player $i$’s cursed plausible conjecture about nature’s initial move is
\[
\bar{\sigma}_\lambda^{I_i}(\theta | h_\emptyset)
= \nu[\sigma](\{((h, \tilde{\theta}), \theta) : (h, \tilde{\theta}) \in P(h_\emptyset)\} | I_i \cap P(h_\emptyset))
= \nu[\sigma](\theta | I_i),
\]
which corresponds to the Bayesian conjecture. The second equality holds because nature’s information set is a singleton. Furthermore, in the framework of multistage games with observed actions, the Bayesian conjecture reduces to
\begin{align*}
\nu[\sigma](\theta | I_i) & = \frac{F\left(\theta_{-i}, \theta_i\right)\left[\prod_{j \neq i} \prod_{1 \leq l \leq t} \sigma_j\left(a_j^l | h^{l-1}, \theta_j\right)\right]}{\sum_{\theta_i^{\prime} \in \Theta_{-i}} F\left(\theta_{-i}^{\prime}, \theta_i\right)\left[\prod_{j \neq i} \prod_{1 \leq l \leq t} \sigma_j\left(a_j^l | h^{l-1},  \theta_j^{\prime}\right)\right]} \\
& = \frac{F\left(\theta_{-i} | \theta_i\right)\left[\prod_{j \neq i} \prod_{1 \leq l \leq t} \sigma_j\left(a_j^l | h^{l-1}, \theta_j\right)\right]}{\sum_{\theta_i^{\prime} \in \Theta_{-i}} F\left(\theta_{-i}^{\prime} | \theta_i\right)\left[\prod_{j \neq i} \prod_{1 \leq l \leq t} \sigma_j\left(a_j^l | h^{l-1}, \theta_j^{\prime}\right)\right]}. \;\;\square
\end{align*}

\subsection*{Proof of Proposition \ref{lemma:phc}}

We first prove the ``only if'' part by contrapositive. Suppose that there exists a non-terminal information set $(h^t, \theta_i)$ such that
$P(h^t, \theta_i) \neq \{(h^t, \theta') : \theta' \in \Theta\}.$
Then, by the definition of the coarsest valid partition, it follows that
$P(h^t, \theta_i) \supsetneq \{(h^t, \theta') : \theta' \in \Theta\}.$
In games with perfect recall, this implies that there exist some $\hat{h}^{t} \in \mathcal{H}^t$ with $\hat{h}^{t} \neq h^t$ and some $\tilde{\theta} \in \Theta$ such that
$(\hat{h}^{t}, \tilde{\theta}) \in P(h^t, \theta_i).$
That is, we have two distinct public histories, $ h^t  \neq \hat{h}^{t}$, such that
$P(h^t, \tilde{\theta}_i) \cap P(\hat{h}^{t}, \tilde{\theta}_i) \neq \emptyset,$
which violates PHC. 
Lastly, the ``if'' part follows immediately from the definition of PHC. This completes the proof. $\square$

\subsection*{Proof of Proposition \ref{claim:ci_cse}}

Let $\Theta = \{(\bar{\theta}_1, \ldots, \bar{\theta}_n) \}$ and consider any behavioral
strategy profile. Because $|\Theta|=1$, for any player $i$ and public history $h^{t-1}$,
$\mu^\chi(\bar{\theta}_{-i}|h^{t-1}, \bar{\theta}_i)=1$. Consequently, the
average behavioral strategy profile of $-i$ is simply:
$$\bar{\sigma}_{-i}(a_{-i}^t|h^{t-1}, \bar{\theta}_i) = 
\mu(\bar{\theta}_{-i}|h^{t-1}, \bar{\theta}_i)
\sigma_{-i}(a_{-i}^t|h^{t-1}, \bar{\theta}_{-i}) =\sigma_{-i}(a_{-i}^t|h^{t-1},  \bar{\theta}_{-i}) $$
for any $a_{-i}^t \in A_{-i}(h^{t-1})$,
suggesting $\sigma_{-i}^\chi(a_{-i}^t|h^{t-1}, \bar{\theta}) = 
\sigma_{-i}(a_{-i}^t|h^{t-1}, \bar{\theta}_{-i})$ for any $\chi\in[0,1]$.
Since the perception about others' strategy profile always aligns with the true strategy profile,
$\chi$-CSE is equivalent to sequential equilibrium for any $\chi\in[0,1]$. $\square$

\subsection*{Proof of Proposition \ref{claim:label_claim}}

If not, there is some $P(h^{t-1}, \theta_i)$ which contains two public histories $h, h' \in \mathcal{H}^t$ where $h\neq h'$.
However, because the game is scrambled, $A_i(h) \neq A_i(h')$.
Therefore, $h$ and $h'$ cannot belong to the same cell of a partition under any partition, which yields a contradiction. $\square$

\subsection*{Formal Analysis of the Simple Signaling Game}

\begin{proposition}\label{claim:signaling_CSE} 
In the simple signaling game in Section \ref{subsec:public_history}, there are two pure $\chi$-CSE:
\begin{itemize}
\item[1.] $[(A,A); (L,R)]$ is a pooling $\chi$-CSE for any $\chi\in[0,1]$.
\item[2.] $[(B,B); (R,R)]$ is a pooling $\chi$-CSE if and only if $\chi
\leq8/9$.
\end{itemize}
\end{proposition}

\noindent\emph{Proof.} We first observe that it is strictly optimal for player~2 to choose $R$ at the information set following player~1’s choice of $B$, whereas at the information set following player~1’s choice of $A$, it is optimal for player~2 to choose $L$ if and only if the belief that player~1 is of type $\theta_1$ satisfies:
$$2\mu(\theta _{1}|A)+[1-\mu(\theta _{1}|A)]\geq 4\mu(\theta
_{1}|A)\iff \mu(\theta _{1}|A)\leq 1/3.$$

\noindent\textbf{Equilibrium 1.}
If both types of player 1 choose $A$, then $\mu(\theta _{1}|A)=1/4$, so
it is optimal for player 2 to choose $L$. Given $a(A)=L$ and $a(B)=R$, it is
optimal for both types of player 1 to pick $A$ since $2>1$. Hence, $[(A,A); (L,R)]$
is a pooling $\chi $-CSE for
any $\chi\in[0,1]$.

\bigskip

\noindent\textbf{Equilibrium 2.}
To support $m(\theta _{1})=m(\theta _{2})=B$ as an equilibrium,
player 2 must choose $R$ at the off-path information set $A,$ which is
optimal if and only if $\mu(\theta _{1}|A)\geq 1/3$. In addition, by
the $\chi$-dampened updating property (see Proposition 3 of \citealp{fong2023cursed}), we know that in a $\chi $-CSE, the belief
system satisfies
\begin{equation*}
\mu(\theta _{2}|A)\geq \frac{3}{4}\chi \iff \mu(\theta _{1}|A)\leq
1-\frac{3}{4}\chi .
\end{equation*}%
Therefore, the belief system must satisfy $\mu(\theta_{1}| A)\in \left[\frac{1}{3},,1-\frac{3}{4}\chi\right]$, requiring $\chi \leq 8/9$. $\square$

\bigskip

\begin{proposition}\label{claim:signaling_sce_wo} 
In the simple signaling game in Section \ref{subsec:public_history}, there are four pure $\phi$-SCE:
\begin{itemize}
\item[1.] $[(A,A); (L,R)]$ is a $\phi$-SCE if and only if
$\phi \leq1/3$.

\item[2.] $[(B,B); (R,R)]$ is a $\phi$-SCE for any $\phi\in[0,1]$.

\item[3.] $[(B,B); (L,R)]$ is a $\phi$-SCE if and only if
$\phi \geq1/3$.

\item[4.] $[(B,A); (L,R)]$ is a $\phi$-SCE if and only if
$\phi = 1/3$.

\end{itemize}
\end{proposition}

\noindent\emph{Proof.} First, as discussed in the proof of Proposition \ref{claim:signaling_CSE}, it is strictly dominant for player 2 to choose $R$ at the information set following player 1's choice of $B$. At the information set following player 1's choice of $A$, it is optimal for player 2 to choose $L$ if and only if the induced belief that player 1 is of type $\theta_1$ satisfies
$\mu^{\phi\text{-SCE}}(\theta_1 | A) \leq 1/3.$ By 
Proposition \ref{lemma_sce_belief}, we can find that:
\begin{enumerate}
\item if $m(\theta_1) = m(\theta_2) = A$, then 
$\mu^{\phi\text{-SCE}}(\theta_1 | A) = 1/4$ and it is optimal 
for player 2 to choose $L$ at the information set following player 1's 
choice of $A$;
\item if $m(\theta_1) = m(\theta_2) = B$, then $\mu^{\phi\text{-SCE}}(\theta_1 | A)\in[0,1]$ and hence both $a(A) = R$ and $L$ can be supported as a 
best response;
\item if $m(\theta_1) = A$ and $m(\theta_2) = B$, then 
$\mu^{\phi\text{-SCE}}(\theta_1 | A) = 1$ and it is optimal for 
player 2 to choose $R$ at the information set following player 1's 
choice of $A$; and 
\item if $m(\theta_1) = B$ and $m(\theta_2) = A$, then 
$\mu^{\phi\text{-SCE}}(\theta_1 | A) = 0$, and it is optimal for 
player 2 to choose $L$ at the information set following player 1's 
choice of $A$.
\end{enumerate}

Second, in the coarsest valid partition of the game, player 2's coarsened information set is $P_2 = \{ \theta_1A, \theta_1B, \theta_2A, \theta_2B\}$, which includes both public histories $A$ and $B$. Consequently, a cursed player 1 in SCE would incorrectly believe that player 2 will behave the same regardless of which message is sent. Therefore,
in a $\phi$-SCE, player 1 conjectures that, with probability $\phi$, player 2's off-path strategy coincides with her on-path strategy, and with probability $1-\phi$, it follows her true strategy.
Because both types of player 1 have the same payoff function, we can conclude the following.
\begin{enumerate}
\item If $a(A) = a(B) = R$, then it is optimal for both types of 
player 1 to choose $B$, implying that 
$[(B,B);(R,R)]$ is a $\phi$-SCE for any $\phi \in [0,1]$, as player 2 chooses the same action on and off the 
path, so the cursedness of SCE does not bind.
\item If $a(A) = L$ and $a(B) = R$, then it is optimal for both types of player 1 to choose
$A$ if and only if
$$2 \geq 4\phi + (1 - \phi) \iff \phi \leq 1/3.$$
Alternatively, it is optimal for both types of player 1 to choose $B$ if and only if
$$1 \geq -\phi + 2(1 - \phi) \iff \phi \geq 1/3.$$
Therefore, $[(A,A);(L,R)]$ is a $\phi$-SCE when $\phi \leq 1/3$, and $[(B,B);(L,R)]$ is a $\phi$-SCE when $\phi \geq 1/3$. Moreover, $[(B,A);(L,R)]$ is also a $\phi$-SCE when $\phi = 1/3$. Note that $[(A,B);(L,R)]$ cannot be supported as a $\phi$-SCE, since player 2 would deviate to choose $R$ upon observing player 1 choosing $A$. $\square$
\end{enumerate}

\subsection*{Formal Analysis of the Crisis-Bargaining Game}

\begin{proposition}\label{prop:crisis_bargaining_cse}
    The unique $\chi$-CSE of the crisis-bargaining game in Section \ref{subsec:cursedness_notion} and \ref{subsec:public_history} is characterized as follows.
    \begin{itemize}
        \item Type $\theta_s$ player 1 chooses T at stage 1 and A at stage 3 with probability 1.
        \item There exists a cutoff value $\hat{\chi} \in (0,1)$ that solves the equation $f(\hat{\chi}) = 0$, where $f(x) \equiv 2x^3 - 4x^2 - 3x + 2$, such that
        \begin{itemize}
            \item Type $\theta_w$ player 1 chooses T with probability $\sigma_{1W}^\chi$ at stage 1 and
            chooses A with probability 0 at stage 3, where 
            \begin{align*}
                \sigma_{1W}^\chi = 
                \begin{cases}
                    \frac{2\chi^{3}-4\chi^{2}+4\chi+1-(1-\chi)\sqrt{16\chi^{2}+8\chi+9}}{2\chi^{3}-4\chi^{2}+3\chi-4}\phantom{0} \qquad \mbox{if} \;\; \chi < \hat{\chi} \\
                    0\phantom{\frac{2\chi^{3}-4\chi^{2}+4\chi+1-(1-\chi)\sqrt{16\chi^{2}+8\chi+9}}{2\chi^{3}-4\chi^{2}+3\chi-4}} \qquad \mbox{if} \;\; \chi > \hat{\chi}.
                \end{cases}
            \end{align*}
            \item Player 2 chooses E with probability $\sigma_2^\chi$ at stage 2, where
            \begin{align*}
                \sigma_2^\chi = 
                \begin{cases}
                    \frac{1}{2}\phantom{1} \qquad \mbox{if} \;\; \chi < \hat{\chi} \\
                    1\phantom{\frac{1}{2}} \qquad \mbox{if} \;\; \chi > \hat{\chi}.
                \end{cases}
            \end{align*}
        \end{itemize}
    \end{itemize}
\end{proposition}

\noindent\emph{Proof.} We first note that at stage 3, it is strictly dominant for
type $\theta_s$ player 1 to choose \emph{A} and for type $\theta_w$ player 1
to choose \emph{BD}. Given this observation, we can conclude that it is strictly dominant 
for type $\theta_s$ player 1 to choose \emph{T} at stage 1, since choosing \emph{T}
yields a strictly positive expected payoff. 
By contrast, type $\theta_w$ player~1 of obtains a positive expected payoff from 
choosing \emph{T} at stage~1 if and only if the probability that player~2 
chooses \emph{C} exceeds $\tfrac{1}{2}$.

Let $\sigma^\chi_{1W}$ denote the probability that player~1 of type $\theta_w$ chooses \emph{T} at stage~1, and let $\mu^\chi$ denote player~2’s posterior belief at stage~2 that player~1 is of type $\theta_s$. In a $\chi$-CSE, player~2 updates their belief using the $\chi$-cursed Bayes’ rule (Proposition~\ref{lemma:rearrange}), implying that
\begin{equation}
    \mu^\chi = \frac{1}{2}\chi + \frac{1}{1+\sigma_{1W}^\chi}(1-\chi).\label{eq:cursed_belief_cse}
\end{equation}
Furthermore, at stage 2, $\chi$-cursed player~2 believes that player~1's future action follows a $\chi$-weighted average of the average behavioral strategy and the true, type-contingent strategy.
In particular, at stage~2, player~2 believes that,
\begin{itemize}
    \item type~$\theta_s$ player 1 will choose \emph{A} at stage 3 with probability $\chi\cdot\mu^\chi + (1-\chi)\cdot 1$;
    \item type~$\theta_s$ player 1 will choose \emph{BD} at stage 3 with probability $\chi\cdot(1-\mu^\chi)+ (1-\chi)\cdot 0$;
    \item type~$\theta_w$ player 1 will choose \emph{A} at stage 3 with probability $\chi\cdot\mu^\chi+ (1-\chi)\cdot 0$;  
    \item type~$\theta_w$ player 1 will choose \emph{BD} at stage 3 with probability $\chi\cdot(1-\mu^\chi) + (1-\chi)\cdot 1$.
\end{itemize}
Thus, $\chi$-cursed player 2's expected payoff of choosing \emph{E}, denoted by $\Pi_2^\chi(E)$, is
\begin{align}
\Pi_2^\phi(E) &= -2\underbrace{\mu^\chi(\chi\mu^\chi+1-\chi)}_{\theta_s, \mbox{ A}}  
+ \underbrace{\mu^\chi\chi(1-\mu^\chi)}_{\theta_s, \mbox{ BD}} + 
2\underbrace{(1-\mu^\chi)\chi\mu^\chi}_{\theta_w, \mbox{ A}} +
\underbrace{(1-\mu^\chi)[\chi(1-\mu^\chi) + (1-\chi)]}_{\theta_w, \mbox{ BD}} \notag \\
&= -2\mu^\chi(\chi\mu^\chi+1-\chi) + 2(1-\mu^\chi)\chi\mu^\chi + (1-\mu^\chi).
\label{eq:In_expected_payoff_cse}
\end{align}
In what follows, we separate the analysis into three cases.

\bigskip

\noindent\textbf{Case 1} [$\sigma_{1W}^\chi = 1$]: In this case, $\mu^\chi=\tfrac{1}{2}$, and
player~2’s expected payoff from choosing \emph{E} at stage~2 is 
$\Pi_2^\chi(E)=\chi-\tfrac{1}{2} > -1$, the payoff from choosing \emph{C} for any $\chi \in [0,1]$.
However, if player~2 chooses \emph{E} with probability one, it is profitable for type $\theta_w$ player~1 to choose \emph{SQ} at stage 1 (i.e., $\sigma_{1W}^\chi=0$).
Therefore, $\sigma_{1W}^\chi=1$ cannot be supported in a $\chi$-CSE.



\bigskip

\noindent\textbf{Case 2} [$\sigma_{1W}^\chi = 0$]: In this case, $\mu^\chi= 1- \tfrac{\chi}{2}$, and player~2’s expected payoff from choosing \emph{E} at stage~2 is $\Pi_2^\chi(E) = -\chi^{3}+2\chi^{2}+\tfrac{3}{2}\chi-2.$
    Since the payoff from choosing \emph{C} at stage~2 is $-1$, player~2 will choose \emph{E} whenever $\Pi_2^\chi(E)>-1$, which is equivalent to 
    $$f(\chi) \equiv 2\chi^3-4\chi^2-3\chi+2 < 0.$$
    The polynomial $f(\chi)$ has a unique root in $(0,1)$, denoted by $\hat{\chi}$, and is strictly decreasing on $(0,1)$. Therefore, for $\chi > \hat{\chi}$, $f(\chi) < 0$ and $\sigma_{1W}^\chi = 0$ can be supported in a $\chi$-CSE. Conversely, for $\chi < \hat{\chi}$, because $f(\chi) > 0$, player~2 will choose \emph{C}, and type $\theta_w$ player~1 has a profitable deviation of $\sigma_{1W}^\chi = 1$.

\bigskip

\noindent\textbf{Case 3} [$\sigma_{1W}^\chi \in (0, 1)$]: To support a $\chi$-CSE in which
type $\theta_w$ player~1  mixes between \emph{T} and \emph{SQ} at stage~1, (1) player~2 must mix between \emph{E} and \emph{C} with equal probability, and (2) type $\theta_w$ player~1 must choose $\sigma_{1W}^\chi$ such that player~2 is indifferent between \emph{E} and \emph{C}.
From the previous analysis, we know that player~2 is indifferent when $\Pi_2^\chi(E) = -1$. By substituting Equation~\eqref{eq:cursed_belief_cse} into Equation~\eqref{eq:In_expected_payoff_cse}, we can simplify $\Pi_2^\chi(E) = -1$ to obtain a quadratic equation in $\sigma_1^\chi$:
    \begin{equation}
        A_\chi({\sigma_{1W}^\chi})^2 + B_\chi(\sigma_{1W}^\chi) + C_\chi = 0
        \label{eq:player_2_indifferent_cse}
    \end{equation}
where $A_\chi = -2\chi^3 + 4\chi^2 - 3\chi + 4 $, 
$B_\chi= 4\chi^3 - 8\chi^2 + 8\chi + 2$ and $C_\chi = -2\chi^3 + 4\chi^2 +3\chi -2$. 
    Therefore, the two roots of Equation \eqref{eq:player_2_indifferent_cse} are
    \begin{equation*}
        \frac{-B_\chi\pm\sqrt{B_\chi^2 - 4A_\chi C_\chi}}{2A_\chi} = \frac{2\chi^{3}-4\chi^{2}+4\chi+1 \pm (1-\chi)\sqrt{16\chi^{2}+8\chi+9}}{2\chi^{3}-4\chi^{2}+3\chi-4}.
    \end{equation*}
    Since $\sigma_{1W}^\chi \in (0,1)$, the only admissible root is the one with the minus sign. 
    That is,  
    $$\sigma_{1W}^\chi = \frac{2\chi^{3}-4\chi^{2}+4\chi+1 - (1-\chi)\sqrt{16\chi^{2}+8\chi+9}}{2\chi^{3}-4\chi^{2}+3\chi-4},$$ 
    which lies strictly between 0 and 1 for all $\chi < \hat{\chi}$. $\square$


\bigskip

\begin{proposition}\label{prop:crisis_bargaining_sce}
    The unique $\phi$-SCE of the crisis-bargaining game in Section \ref{subsec:cursedness_notion} and \ref{subsec:public_history} can be characterized as follows.
    \begin{itemize}
        \item Type $\theta_s$ player 1 chooses T at stage 1 and A at stage 3 with probability 1.
        \item Type $\theta_w$ player 1 chooses T with probability $\sigma_{1W}^\phi$ at stage 1 and A with probability 0 at stage 3, where 
        $$\sigma_{1W}^\phi = \frac{1-4\phi+\sqrt{16\phi^2+8\phi+9}}{4}.$$
        \item Player 2 chooses E with probability $\tfrac{1}{2}$ at stage 2.
    \end{itemize}
\end{proposition}

\noindent\emph{Proof.} Applying the same argument as in the proof of Proposition~\ref{prop:crisis_bargaining_cse}, it follows that choosing \emph{T} at stage~1 and 
\emph{A} at stage~3 is optimal for type $\theta_s$ player~1, whereas
choosing \emph{BD} at stage~3 is optimal for type $\theta_w$ player~1.
Let $\mu^\phi$ denote player 2's posterior belief at stage 2 that player 1 is of type $\theta_s$. 
    Given $\sigma_{1W}^\phi$, the probability of type $\theta_w$ player 1 choosing \emph{T} at stage 1, Proposition \ref{lemma_sce_belief} implies 
    \begin{equation}
        \mu^\phi = \frac{1}{1+\sigma_{1W}^\phi}.
        \label{eq:cursed_belief_sce}
    \end{equation}
At stage~2, player~2 believes that, 
\begin{itemize}
    \item type~$\theta_s$ player 1 will choose \emph{A} at stage 3 with probability $\phi\cdot\mu^\phi + (1-\phi)\cdot 1$;
    \item type~$\theta_s$ player 1 will choose \emph{BD} at stage 3 with probability $\phi\cdot(1-\mu^\phi)+ (1-\phi)\cdot 0$;
    \item type~$\theta_w$ player 1 will choose \emph{A} at stage 3 with probability $\phi\cdot\mu^\phi+ (1-\phi)\cdot 0$;  
    \item type~$\theta_w$ player 1 will choose \emph{BD} at stage 3 with probability $\phi\cdot(1-\mu^\phi) + (1-\phi)\cdot 1$.
\end{itemize}
Therefore, $\phi$-cursed player 2's expected payoff of choosing \emph{E}, denoted by $\Pi_2^\phi(E)$, is
\begin{align}
\Pi_2^\phi(E) &= -2\underbrace{\mu^\phi(\phi\mu^\phi+1-\phi)}_{\theta_s, \mbox{ A}}  
+ \underbrace{\mu^\phi\phi(1-\mu^\phi)}_{\theta_s, \mbox{ BD}} + 
2\underbrace{(1-\mu^\phi)\phi\mu^\phi}_{\theta_w, \mbox{ A}} +
\underbrace{(1-\mu^\phi)[\phi(1-\mu^\phi) + (1-\phi)]}_{\theta_w, \mbox{ BD}} \notag \\
&= -2\mu^\phi(\phi\mu^\phi+1-\phi) + 2(1-\mu^\phi)\phi\mu^\phi + (1-\mu^\phi).
\label{eq:In_expected_payoff_sce}
\end{align}
In what follows, we separate the analysis into three cases.

\bigskip

\noindent\textbf{Case 1} [$\sigma_{1W}^\phi = 1$]: In this case, $\mu^\phi=\tfrac{1}{2}$, implying that player~2’s expected payoff from choosing \emph{E} at stage~2 is $\Pi_2^\phi(E)=\phi-\tfrac{1}{2}$.
    Applying the same argument as in the proof of Proposition \ref{prop:crisis_bargaining_cse} for case 1, it follows that $\sigma_{1W}^\phi=1$ cannot be supported in a $\phi$-SCE.

\bigskip

\noindent\textbf{Case 2} [$\sigma_{1W}^\phi = 0$]: In this case, $\mu^\phi=1$, implying that player~2’s expected payoff from choosing \emph{E} at stage~2 is $\Pi_2^\phi(E)= -2$, which is strictly less than the payoff from choosing \emph{C} for any $\phi \in [0,1]$. 
    However, if player 2 chooses \emph{C} with probability one, it is profitable for type $\theta_w$ player 1 to choose \emph{T} at stage 1 (i.e., $\sigma_{1W}^\phi = 1$). 
    Therefore, $\sigma_{1W}^\phi = 0$ cannot be supported in a $\phi$-SCE.

\bigskip

\noindent\textbf{Case 3} [$\sigma_{1W}^\phi \in (0, 1)$]: At stage 1, type $\theta_w$ player 1 is indifferent between \emph{T} and \emph{SQ} when player 2 chooses \emph{E} with probability $\tfrac{1}{2}$. 
    Player 2, on the other hand, is indifferent between \emph{E} and \emph{C} when $\Pi_2^\phi(E) + 1 = 0$.
    By substituting Equation (\ref{eq:cursed_belief_sce}) into
    Equation (\ref{eq:In_expected_payoff_sce}), we can simplify $\Pi_2^\phi(E) + 1 = 0$ and obtain a quadratic equation in $\sigma_{1W}^\phi$:
    \begin{equation}
        2({\sigma_{1W}^\phi})^2 + (4\phi+1)\sigma_{1W}^\phi - 1 = 0.
        \label{eq:player_2_indifferent_sce}
    \end{equation}
Using the quadratic formula, we obtain that the two roots of Equation \eqref{eq:player_2_indifferent_sce} are
    \begin{equation*}
        \frac{-(4\phi+1)\pm\sqrt{16\phi^2+8\phi+9}}{4}.
    \end{equation*}
    Given that $\sigma_{1W}^\phi \in (0,1)$, only the root corresponding to the positive sign in the numerator is admissible for $\phi \in [0,1]$.
    Specifically,
    $$\sigma_{1W}^\phi =\frac{-(4\phi+1)+\sqrt{16\phi^2+8\phi+9}}{4},$$ 
    which lies strictly between 0 and 1 for all $\phi \in [0,1]$. $\square$

\end{document}